\documentclass{article}

    \usepackage[preprint]{neurips_2024}

\usepackage[utf8]{inputenc} 
\usepackage[T1]{fontenc}    
\usepackage{hyperref}       
\usepackage{url}            
\usepackage{booktabs}       
\usepackage{amsfonts}       
\usepackage{nicefrac}       
\usepackage{microtype}      
\usepackage{xcolor}         
\usepackage{graphicx}
\usepackage[utf8]{inputenc} 
\usepackage[T1]{fontenc}    
\usepackage{hyperref}       
\usepackage{url}            
\usepackage{booktabs}       
\usepackage{amsfonts}       
\usepackage{nicefrac}       
\usepackage{microtype}      
\usepackage{natbib}
\setcitestyle{numbers}
\usepackage{subfig}
\usepackage{bbding}
\usepackage{caption}

\usepackage{algorithm}
\usepackage{algpseudocode}
\usepackage{graphicx} 
\usepackage{amsmath}
\usepackage{amssymb}
\usepackage{mathrsfs}
\usepackage{stmaryrd}
\usepackage{wrapfig}
\usepackage{graphicx}
\newcommand*{\img}[1]{%
    \raisebox{-.1\baselineskip}{%
        \includegraphics[
        height=\baselineskip,
        width=\baselineskip,
        keepaspectratio,
        ]{#1}%
    }%
}
\newcommand*{\lowerimg}[1]{%
    \raisebox{-.3\baselineskip}{%
        \includegraphics[
        height=\baselineskip,
        width=\baselineskip,
        keepaspectratio,
        ]{#1}%
    }%
}
\newtheorem{definition}{Definition}

\title{Representation Learning For Efficient Deep Multi-Agent Reinforcement Learning}

\author{%
Dom Huh$^{1}$ \quad Prasant Mohapatra$^{1,2}$ \\
$^1$UC Davis \quad $^2$University of South Florida\\
\texttt{dhuh@ucdavis.edu$^1$}\\
\texttt{pmohapatra@usf.edu$^2$}
}

\begin{document}
\setcitestyle{square}

\maketitle

\begin{abstract}
Sample efficiency remains a key challenge in multi-agent reinforcement learning (MARL). A promising approach is to learn a meaningful latent representation space through auxiliary learning objectives alongside the MARL objective to aid in learning a successful control policy. In our work, we present MAPO-LSO (Multi-Agent Policy Optimization with Latent Space Optimization) which applies a form of comprehensive representation learning devised to supplement MARL training. Specifically, MAPO-LSO proposes a multi-agent extension of transition dynamics reconstruction and self-predictive learning that constructs a latent state optimization scheme that can be trivially extended to current state-of-the-art MARL algorithms. Empirical results demonstrate MAPO-LSO to show notable improvements in sample efficiency and learning performance compared to its vanilla MARL counterpart without any additional MARL hyperparameter tuning on a diverse suite of MARL tasks.
\end{abstract}

\section{Introduction}
A multi-agent control system consists of multiple decision-making entities within a shared environment, each tasked with achieving some objectives defined by a reward signal. Multi-agent reinforcement learning (MARL) offers a learning paradigm that optimizes for emergent rational behaviors within agents through interactions with the environment and one another to achieve an equilibrium \cite{huh2023multiagent}. In recent years, deep MARL has proven successful in numerous domains, including robotics teams \cite{huh2023ist}, networking applications \cite{qu2022scalable}, and various social scenarios that require multi-agent interactions \cite{bucsoniu2010multi}. However, deep reinforcement learning (RL) has historically suffered from sample inefficiency, requiring a costly amount of interaction experience to learn valuable behaviors. This challenge stems largely from the high variance in existing RL algorithms paired with the data-intensive nature of deep neural networks \cite{sutton2018reinforcement}. Unfortunately, MARL applications face additional learning pathologies and complexities \cite{palmer2020thesis} such as exponential computational scaling with respect to the number of agents and the dynamic challenge of equilibrium computation \cite{daskalakis2023complexity}.

To remedy this issue, recent MARL efforts have concentrated on the concept of centralized training and decentralized execution (CTDE) \cite{lowe2017multi, kuba2021trust, yu2022surprising}. In CTDE, agents are trained with access to global state information while retaining autonomy, meaning the agents can make decisions based only on local information during execution. Despite the empirical improvements from CTDE, sample inefficiency remains an elusive challenge. We argue that the CTDE paradigm does not fully address the underlying limitations of RL algorithms, i.e. the sparsity and variance of its learning signals.

A natural solution to address this issue is to curate additional learning signals that supplement and enrich the RL learning process. This approach of imposing further inductive bias has proven effective in prior works at enhancing the training of control policies in single-agent RL \cite{jaderberg2016reinforcement}. The new objectives that are introduced range from reinforcing similarities and dissimilarities within temporal or spatial locality \cite{laskin2020curl, stooke2021decoupling} to instilling information regarding different aspects of the tasks, such as the transition dynamics \cite{schwarzer2020data}, into the latent state space. Importantly, the main takeaway from these efforts is to learn a rich latent state space that understands and is coherent with the task dynamics and itself \cite{ni2024bridging}. However, much of these techniques of representation learning has yet to be fully realized and extended to a MARL context. 

\begin{figure}
    \centering
    \includegraphics[width=0.9\textwidth]{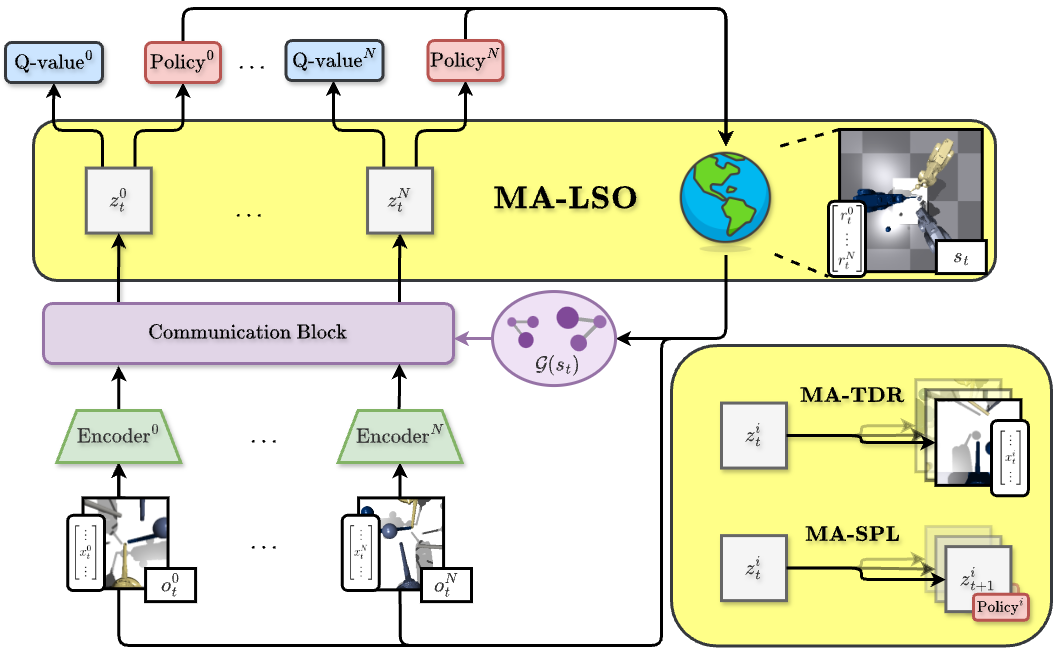}
    \caption{A high-level illustration of the MAPO-LSO framework. For each agent $i = \{0,\dots,N\}$, the encoders (\img{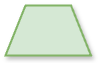}) embed their observations $o_t^i$ and propagates their encodings through a communication block (\img{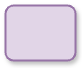}) that is subject to a communication network $\mathcal{G}(s_t)$. Once the agents communicate, the latent state $z_t^i$ (\lowerimg{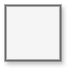}) is computed and used as inputs for its policy (\lowerimg{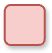}) and value function (\lowerimg{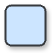}). For our MA-LSO procedure, the latent states are optimized using MA-Transition Dynamics Reconstruction (MA-TDR) and MA-Self-Predictive Learning (MA-SPL). These two learning processes are outlined in Section \ref{ma-lso} and loosely can be thought of as instilling the capability of inferring the observations and the next latent states of all agents from the current latent state.} 
    \label{fig:mapo-lso}
\end{figure}

In this work, we propose MAPO-LSO (Multi-Agent Policy Optimization with Latent Space Optimization), a generalized MARL framework, outlined in Figure \ref{fig:mapo-lso}, that leverages latent space optimization (LSO) in a multi-agent setting under the CTDE framework. Specifically, we show that current state-of-the-art MARL algorithms, such as MAPPO \cite{yu2022surprising}, HAPPO \cite{kuba2021trust}, MASAC \cite{haarnoja2018soft}, and MADDPG \cite{lowe2017multi} benefit from our multi-agent LSO (MA-LSO) learning process with trivial modifications. Our experiments demonstrate significant improvements in not only the sample efficiency but also in the overall performance over $18$ diverse tasks in VMAS \cite{bettini2022vmas} and $5$ robotic team tasks in IsaacTeams \cite{huh2023ist} over all algorithms under fixed model architectures and MARL hyperparameters setting.

Our contributions are as follows:
\begin{itemize}
    \item [1.] We introduce a novel MARL framework, MAPO-LSO, that integrates MA-LSO, a comprehensive form of representation learning into the MARL training. MA-LSO is broken down into two parts: MA-Transition Dynamics Reconstruction and MA-Self-Predictive Learning. Hence, we provide a new perspective on the intuition behind the usage and integration of both learning processes in a multi-agent control setting.
    \item [2. ] We study the application of pretraining, uncertainty-aware modeling techniques for agent-modeling and phasic optimization within our MAPO-LSO framework to improve learning performance, specifically in terms of convergence and stability.
    \item [3.] We extend and experiment using several state-of-the-art MARL algorithms on our MAPO-LSO framework on a variety of tasks with diverse nature of interactions and multi-modal data, presenting further ablation studies on design choices to showcase the improvements of our MAPO-LSO framework.
\end{itemize}

\section{Related Works}
\paragraph{Sample-Efficiency in MARL} A number of recent works have addressed the sample efficiency problem in deep MARL ranging from developing vectorized and parallelizable simulation platforms \cite{huh2023ist, bettini2022vmas}, improving exploration strategies to collect diverse and useful samples \cite{li2023coin}, pre-training on a dataset of demonstrations \cite{qiu2022sample}, utilizing off-policy and/or model-based approaches \cite{liu2023efficient}, and learning on offline datasets \cite{yang2021believe}. While these prior efforts are not necessarily orthogonal to our efforts, the focus of this paper is on introducing a form of multi-agent representation learning that improves how much is learned from each sample by guiding the optimization of the latent state space for MARL tasks.

\paragraph{Representation Learning in MARL} The concept of representation learning has previously been applied in MARL applications through masked observation reconstruction \cite{kim2023sample, song2023ma2cl}, auxiliary task-specific predictions \cite{shang2021agent}, self-predictive learning in joint latent space \cite{feng2022joint}, and contrastive learning on the observation embedding space \cite{hu2024attentionguided}. In our study, our proposed MA-LSO takes a more comprehensive measure by applying two forms of representation learning that enforce consistency between the latent state space and the transition dynamics and within itself as a self-predictive representation space.

\section{Preliminaries}
In this work, we consider an extension of the stochastic game framework \cite{Shapley1953StochasticGame} called the networked Bayesian game \cite{harsanyi1967games, huh2023multiagent}.
\begin{definition} A networked Bayesian game is defined by a tuple $\langle \mathscr{I}, \mathcal{S}, \mathcal{O},\mathcal{A}, \mathcal{T}, R, \omega, \mathcal{G} \rangle$. 
\begin{itemize}
    \item $\mathscr{I} = \{0,...,N\}$ is the set of $N$ agents.
    \item $\mathcal{S}$ is the global state space.
    \item $\mathcal{O} = \prod\limits_{i \in \mathscr{I}} \mathcal{O}^i$ is the joint observation space, where $\mathcal{O}^i$ is the observation space of agent $i$.
    \item $\mathcal{A} = \prod\limits_{i \in \mathscr{I}} \mathcal{A}^i$ is the joint action space, where $\mathcal{O}^i$ is the action space of agent $i$.
    \item $\mathcal{T}: \mathcal{S} \times \mathcal{A} \mapsto P(\mathcal{S})$ is the state transition operator, mapping the state-action space to the probability of the next states.
    \item $R = \prod\limits_{i \in \mathscr{I}} R^i$, is the joint reward function, where $R^i:\mathcal{S} \times \mathcal{A} \mapsto \mathbb{R}$ is the reward function for agent $i$.
    \item $\omega = \prod\limits_{i \in \mathscr{I}} \omega^i$ is the joint type/belief space, where $\omega^i$ is the belief space of agent $i$.
    \item $\mathcal{G}: \mathcal{S} \mapsto \mathscr{I} \times \mathscr{I}$ is the mapping from the state to an adjacency matrix that defines the communication graph between all agents.
\end{itemize}
\end{definition}

We optimize for the Bayes-Nash equilibrium, where each agent $i$ learns a best-response policy $\pi^i: \mathcal{O}^i \times \omega^i \mapsto \mathcal{A}^i$, by maximizing the expected \textit{ex interm} return of individual agent $G^i$. 

\begin{equation}
    \forall i\in \mathscr{I}, G^i = \mathbb{E}_{\tau \sim \mathcal{T}_{\pi^i}}[\sum_{t=0} R^i(s_t, a_t)] \text{ where }\tau = \{s_0,a_0,\dots\}
\end{equation}

\paragraph{Deep Reinforcement Learning} The field of deep RL presents general control optimization algorithms using deep neural network approximations: canonically existing in the form of Q-learning, policy gradient, and actor-critic methods \cite{sutton2018reinforcement}. With Q-learning approaches, the optimal state-value or action-value function $Q^{\pi^*}(s, a)$ is learned, where $Q^\pi(s, a)$ maps the state-action pair to its expected return following a policy $\pi$. Policy gradient methods directly optimize the policy via gradient ascent over the expected return. Actor-critic methods stabilize the policy gradient by approximating the offset reinforcement with a learned Q-function under the current policy. These optimization schemes have been extended to and studied under a multi-agent context, demonstrating promising results \cite{yu2022surprising}.

\paragraph{Latent Space Optimization} Latent space optimization (LSO) is a form of representational learning that is often used in unison with generative modeling \cite{zhou2023deep}, where LSO leverages model-based optimization that learns an approximation of the objective function under a learned latent space \cite{tripp2020sample}. In RL, LSO is often used to map various aspects of the environment model into a latent space to assist with the RL training \cite{hafner2019dream}. In this work, we explore this pretense within a MARL setting. 

\section{Multi-agent Latent Space Optimization}\label{ma-lso}
Our approach, MA-LSO, optimizes a latent state representation $z^i_t \in Z^i_t$ for each agent $i \in \mathscr{I}$ to supplement the ``sample-inefficient" MARL optimization. To achieve this goal, we employ two processes: MA-Transition Dynamics Reconstruction (MA-TDR) and MA-Self-Predictive Learning (MA-SPL). These processes draw inspiration from previous work on single-agent model-based RL \cite{hafner2019dream} and representational learning methods for RL \cite{schwarzer2020data, grill2020bootstrap, guo2020bootstrap}, unifying the concepts of TDR and SPL in a manner that complements one another while considering the multi-agent nature of the task.

\subsection{MA-Transition Dynamics Reconstruction}\label{tdr}
MA-TDR learns an approximation of the transition dynamics of the environment by mapping the underlying ``true" state to a latent state representation space $\mathcal{Z}_t$, grounding $\mathcal{Z}_t$ to the realities of the task's dynamics. To implement this, we make use of recurrent modeling and multi-agent predictive representation learning (MA-PRL).

\begin{wrapfigure}[27]{r}{0.45\textwidth}
    \centering
    \includegraphics[width=0.45\textwidth]{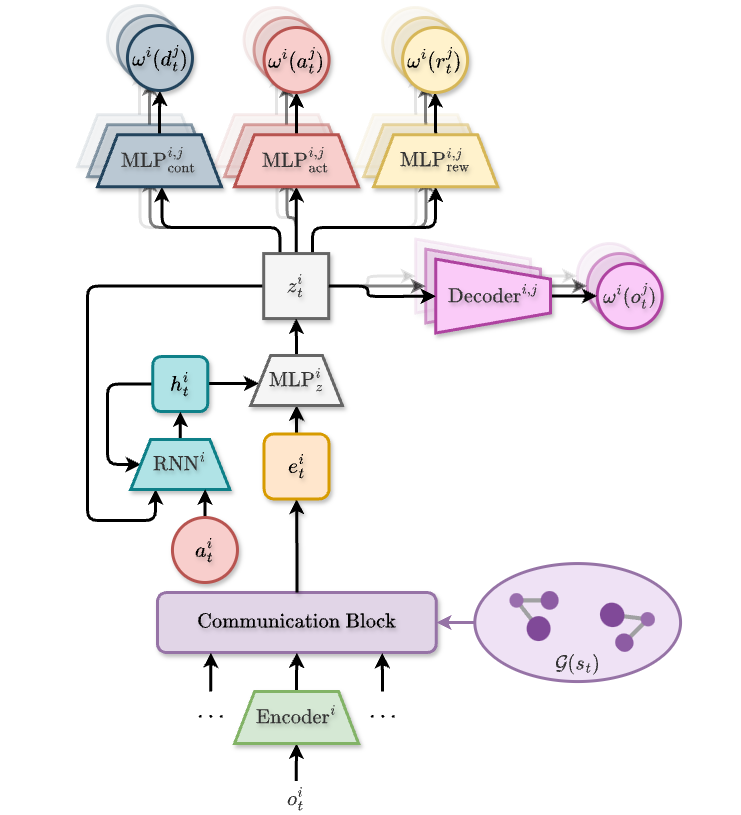}
    \caption{A detailed visualization of the MA-TDR modeling procedure with the auxiliary modules used to reconstruct transition dynamics for recurrent modeling and MA-PRL.}
    \label{fig:mapo-lso}
\end{wrapfigure}

\paragraph{Recurrent Modeling} For each agent $i$, we maintain a recurrent state $h_t^i$ that holds information regarding its history and is realized using a recurrent neural network $\textsc{rnn}$.
$$h^i_t = \textsc{rnn}^i( \llbracket z^i_{t-1}, a^i_{t-1} \rrbracket; h^i_{t-1})$$
From the observation $o_t^i$, an encoder computes an embedding to be passed into a communication block to generate $e_t^i$. The latent state space $\mathcal{Z}^i_t$ is computed using a multi-layer perceptron $\textsc{mlp}$ that processes the embedding $e_t^i$ and the recurrent state $h_t^i$.
$$e^i_t = \text{Communication Block}(\text{Encoder}^{i}(o^i_t) | \mathcal{G}(s_t))$$
$$z^i_t \sim \mathcal{Z}^i_t = P(\textsc{mlp}_z^{i}(\llbracket e^i_t, h^i_t \rrbracket ))$$
where $\mathcal{Z}^i_t$ is a mixture of categorical distributions \cite{hafner2023mastering}. The purpose of this recurrent modeling is to ensure that the latent state is expressive enough such that it is sufficient to recollect information needed for decision-making from the agent's history and can tractably perform the other auxiliary tasks posed in MA-PRL and MA-SPL. 

\paragraph{MA-Predictive Representation Learning} In MA-PRL, we take explicit measures to ensure that the latent state $z_t$ contains information regarding the transition dynamics by reconstructing and inferring various aspects of the transition dynamics -- namely, the observation $o_t$, reward $r_t$ and termination $d_t$ -- from the latent state $z_t$.

Firstly, MA-PRL incorporates CURL \cite{laskin2020curl}, a contrastive learning framework that guides the latent state $z^i_t$ produced by $o^i_t$ to be similar to the $\hat{z}^i_t$ produced by an augmented $\hat{o}^i_t$.

Next, we task each agent $i$ to maintain a belief over its own as well as the other agents' observations, policies, rewards, and termination. These beliefs are computed as a function of their latent state $z_t^i$. To ensure the feasibility of these beliefs, we experiment with Monte-Carlo dropout to address the inherent epistemic uncertainty \cite{gal2016dropout, krishnan2022bayesiantorch}.
$$\omega^i(o^{j}_t) = \text{Decoder}^{i,j}(z^i_t) \quad \omega^i(a^j_t) = \textsc{mlp}^{i,j}_{\text{act}}(z^i_t) \quad \omega^i(r^{j}_t) = \textsc{mlp}^{i,j}_{\text{rew}}(z^i_t) \quad 
\omega^i(c^{j}_t) = \textsc{mlp}^{i,j}_{\text{cont}}(z^i_t)$$
where $c^j_t = (1-d^j_t)$ is the continue signal for agent $j$. In terms of our implementation, we adhere to the same protocols set in \cite{hafner2023mastering}, approximating the reward using a symlog twohot distribution and the continue signal using onehot distribution. Additionally, we temporally-smoothed the reward signals with Gaussian-smoothing to ease the task of reward distribution approximation \cite{lee2024dreamsmooth}.

The combination of the two concepts, recurrent modeling and MA-PRL, makes up the MA-TDR process. The overall loss for MA-TDR is defined as:
\begin{align}
    \mathscr{L}_{tdr} \doteq \mathbb{E}_{(o_t, a_t, r_t, c_t)\sim\mathcal{D}} [\sum_{i,j \in \mathscr{I}} &\underbrace{+\frac{\exp (s(\hat{z}^i_t, z^i_t))}{\sum\limits_{k\in \mathscr{I}} \exp(s(\hat{z}^i_t, z^k_t))}}_{\text{CURL loss}} \underbrace{-\ln P(\omega^i(o^{j}_t)=o^j_t)}_{\text{obs log loss}}  \nonumber\\ &\underbrace{+ \textsc{h}(\omega^i(a^j_t), a^j_t)}_{\text{action loss}} \underbrace{-\ln P(\omega^i(r^{j}_t)=r^{j}_t)}_{\text{reward log loss}} \underbrace{-\ln P(\omega^i(c^{j}_t)=c^{j}_t)}_{\text{continue log loss}} ]
\end{align}\label{equation: tdr}

where $s(\cdot)$ is the similarity measure adopted from \cite{laskin2020curl}, $\mathcal{D}$ is an experience replay buffer, $\textsc{h}(\cdot)$ is Huber loss and $sg(\cdot)$ is the stop-gradient operator. For each loss term, we append a scaling hyperparameter to each loss term to avoid dominating gradients and general performance reasons.

\begin{figure}[b]
    \centering
    \includegraphics[width=\textwidth]{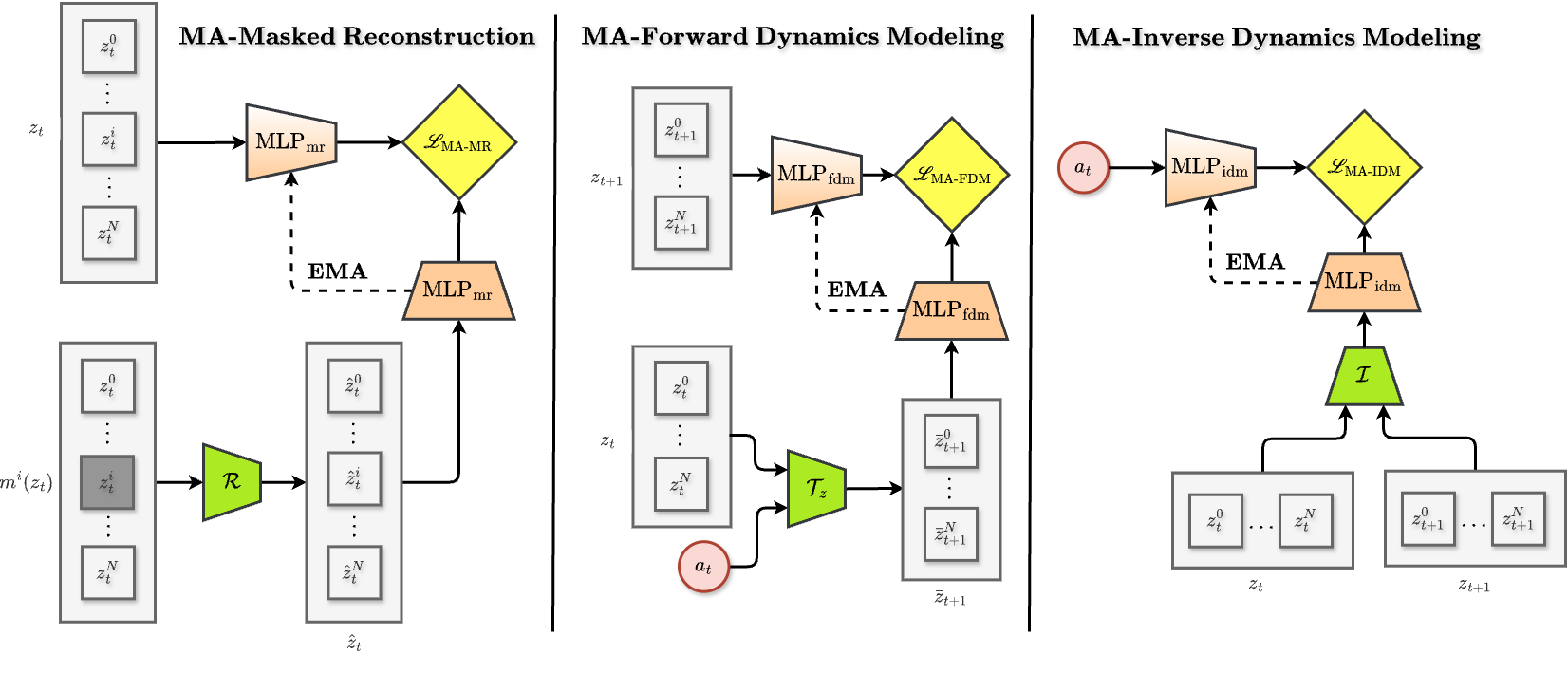}
    \caption{The three MA-SPL subprocesses of MA-MR, MA-FDM and MA-IDM are shown.}
    \label{fig:spl}
\end{figure} 

\subsection{MA-Self-Predictive Learning} \label{maspl}
The desideratum of MA-SPL is to learn a $\mathcal{Z}_t$ that is sufficient to predict the expected $\mathcal{Z}_{t+1}$ \cite{givan2003equivalence, subramanian2022approximate}. Intuitively, the learned latent state space is optimized to be consistent with itself and its own latent dynamics \cite{tang2023understanding}. Moreover, we extend the concept of SPL \cite{schwarzer2020data} to a multi-agent setting, where now, we consider the presence of other agents in the environment and thereby enforce a structural relation \cite{tseng2022offline} and consistency amongst the agents in a centralized manner.

\paragraph{MA-Masked Reconstruction (MA-MR)} Inspired by \cite{kim2023sample, song2023ma2cl}, MA-MR encourages inter-predictive representation between agents' latent space. Similar to \cite{wen2022multi}, MA-MR treats the agents' latent states as a sequence. Concretely, MA-MR utilizes a contrastive learning paradigm such that a masked joint latent state can reconstruct the joint latent state $z_t$. This masking process is applied on the agent-level. Hence, if the latent state of agent $i$ is masked $m_i(z_t)$, the joint latent spaces of the other agents $z^{\mathscr{I}\setminus i}_t$ is sufficient to reconstruct $z^i_t$. In our implementation, we adopt the framework from \cite{song2023ma2cl}, using a self-attentive reconstruction model $\mathcal{R}(\cdot)$ to process the masked latent state as shown in Figure \ref{fig:spl}.
$$\bar{z_t} = \mathcal{R}(m_i(z_t))$$
\paragraph{MA-Forward Dynamics Modeling (MA-FDM)} The objective of MA-FDM is to ensure that the information contained in the current joint latent state $z_t$ and the joint action $a_t$ is sufficient to infer the next joint latent state $z_{t+1}$ \cite{feng2022joint}. To implement this, we define a transition head $\mathcal{T}_z(\cdot)$ which is realized using a cross-attention head \cite{vaswani2017attention} that maps the joint latent state and the joint action to the next joint latent space.
$$\bar{z}_{t+1} = \mathcal{T}_z(z_t, a_t)$$
\paragraph{MA-Inverse Dynamics Modeling (MA-IDM)} MA-IDM aims to achieve the following objective: Given the current and next joint latent state, the joint action that realized that transition from the current to the next joint latent state can be deduced. In our work, we use an inverse head $\mathcal{I}(\cdot)$ which is realized using a self-attentive model that maps the current and next joint latent state to the joint action space.
$$\bar{a}_{t} = \mathcal{I}(z_t, z_{t+1})$$

The overall loss for MA-SPL is defined as:
\begin{align}
    \mathscr{L}_{spl} \doteq \mathbb{E}_{(o_t, a_t, o_{t+1})\sim \mathcal{D}}[\sum_{i\in \mathscr{I}} &\underbrace{\frac{\exp (s(\bar{z}^i_t, z^i_t))}{\sum\limits_{j\in \mathscr{I}} \exp(s(\bar{z}^i_t, z^j_t))}}_{\text{MA-MR Loss}/\mathscr{L}_{\text{MA-MR}}}
    \underbrace{+\textsc{h}(\bar{z}_{t+1}, z_{t+1})}_{\substack{\text{MA-FDM Loss}\\/\mathscr{L}_{\text{MA-FDM}}}} \underbrace{+ \textsc{h}(\bar{a}_t,a_t)}_{\substack{\text{MA-IDM Loss}\\/\mathscr{L}_{\text{MA-IDM}}}}]
\end{align}

\paragraph{MLP Heads} Following recent works on contrastive learning frameworks \cite{chen2003improved, chen2020simple}, we introduce MLP projection heads for CURL, MA-MR, MA-FDM, and MA-IDM learning processes. Moreover, we adopt a momentum-like update similar to these prior efforts. This addition is shown in Figure \ref{fig:spl} for MA-MR, MA-FDM, and MA-IDM.

\subsection{Integrating MA-LSO to Multi-agent Policy Optimization}
Our proposed approach, MA-LSO, can easily be appended to popular MARL algorithms with minor adjustments to form MAPO-LSO. The central challenge is the use of recurrent modeling, which raises several implementation challenges for some algorithms \cite{kapturowski2018recurrent} involving adjustments in the experience replay buffer $\mathcal{D}$ and maintenance of the recurrent state. Otherwise, appending the MA-LSO learning process is trivial and can be performed concurrently with the MARL training.

\paragraph{On-policy MARL} In general, we define a shared replay buffer $\mathcal{D}$ that we sample batches of transitions from to compute both MA-LSO and MARL objectives. However, for on-policy MARL algorithms that cannot learn on the offline data, we found that learning on offline data during the MA-LSO process is necessary to promote good generalization and stable learning by learning on a diverse dataset. Therefore, we ensure that we maintain both on-policy and off-policy data in $\mathcal{D}$ such that online data is used for the MARL training but off-policy data is still available for the MA-LSO process.

\paragraph{Phasic Optimization}\label{phasicdescr} For certain MARL algorithms, notably MADDPG and MASAC, we chose to follow the training methodology outlined in \cite{fujimoto2018addressing}. This involved the utilization of target networks and delayed policy updates. These techniques are intended to mitigate the learning variance and enhance overall performance. However, despite these efforts, we still observed that the training remained too sensitive to hyperparameters, likely due to the use of a model architecture that shares parameters between the policy and value function (i.e. the encoder) and the phasic nature of learning.

To mitigate this instability, we recognized the need to incorporate a phasic regularization term inspired by the work of \cite{cobbe2021phasic}. This regularization term constrains the policy divergence during all non-policy updates and thereby promotes a more stable learning environment. For HAPPO, we also enforce this regularization term during the sequential policy updates such that the shared encoder, which exists within the centralized critic, does not diverge from the other agents' behaviors that are not being updated. In our study, instead of using a KL divergence, we use Huber loss to constrain the divergence of actions (i.e. of the policies) utilizing the old and new encoders.

\paragraph{Pre-training} The MA-LSO objective can be used as a pre-training paradigm similar to \cite{schwarzer2021pretraining}, where $\mathcal{D}$ is pre-filled with an exploratory/random policy of transitions and is trained on $\mathscr{L}_{tdr}$ and $\mathscr{L}_{spl}$.

\section{Experiments}
For our experiments, we use the tasks in Vectorized Multi-agent Simulator (VMAS) tasks and IsaacTeams (IST) to provide a comprehensive evaluation of a diverse collection of multi-agent tasks, selecting $18$ diverse tasks from VMAS and $5$ tasks from IST as shown in Appendix \ref{alltasks}.

\paragraph{Experimental Setup} The scenarios parameters for VMAS and IST environments are taken from prior works \cite{bettini2022vmas, Bettini_BenchMARL_Benchmarking_Multi-Agent_2023, huh2023ist}. The four MARL algorithms chosen for our experiments are MAPPO \cite{yu2022surprising}, HAPPO \cite{kuba2021trust}, MASAC \cite{haarnoja2018soft}, and MADDPG \cite{lowe2017multi}; all of which are considered competitive MARL baselines. For all experiments, the MARL hyperparameters are initially tuned using a random search for the vanilla MARL algorithms (i.e. without MA-LSO), then kept fixed and trained with our MAPO-LSO method for that specific task. Further implementation details are provided in Appendix \ref{appendix:model} and \ref{hyperparameters}. All experiments presented in this work were executed on $3$ Nvidia RTX A6000 and Intel Xeon Silver 4214R @ 2.4GHz.

\subsection{Results} 
In this section, we evaluate the overall performance and sample efficiency of MAPO-LSO paired with popular MARL algorithms. Here, performance refers to the collective return achieved and the sample efficiency is measured by the performance with respect to the number of data samples used, meaning the better the performance at a given number of environment transitions learned on, the higher the sample efficiency. We additionally conduct further ablation studies to investigate and analyze each component of our MAPO-LSO method and study if any other improvements or degradations are realized at a more granular level.

To provide a concise comparison against our method, we present much of our results in a normalized scale. This involves aggregating and scaling the results from each experiment, algorithm, and task \cite{colas2018many, gorsane2022towards}.

\begin{figure}[b]
    \centering
    \subfloat[MAPO-LSO\label{normalized-efficacy}]{\includegraphics[width=0.33\textwidth]{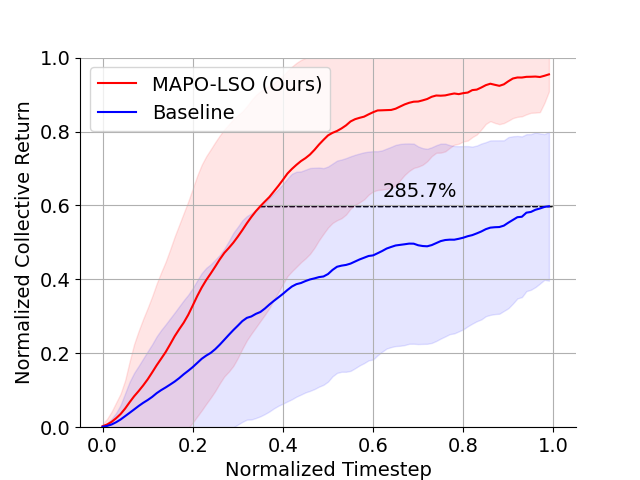}}
    \subfloat[Phasic Regularization\label{normalized-phasic}]{\includegraphics[width=0.33\textwidth]{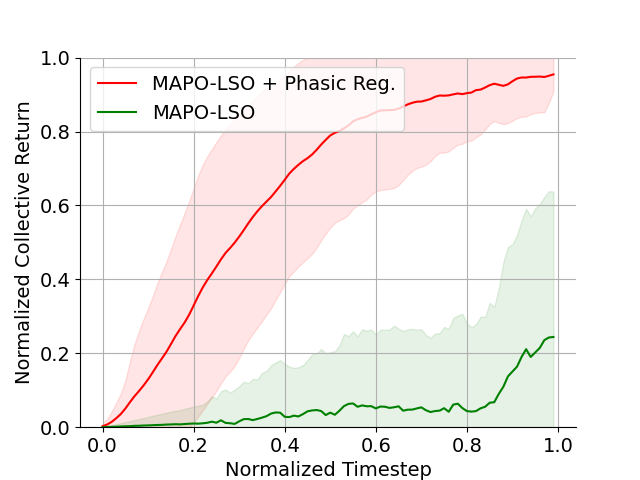}}
    \subfloat[Uncertainty Modeling\label{normalized-num}]{\includegraphics[width=0.33\textwidth]{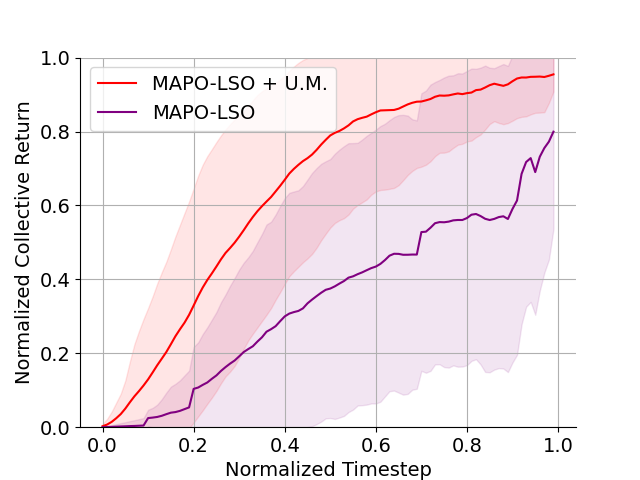}}
    \caption{The graphs compare the collective returns under a normalized scale between various components introduced in this work --- namely,  MAPO-LSO, phasic regularization, and uncertainty modeling (U.M.) --- over all VMAS and IST tasks and MARL algorithms, except for Figure \ref{normalized-phasic}, which normalizes over HAPPO, MADDPG and MASAC. The error bars indicate $\pm1$ std deviations. The results for the individual runs of all experiments are provided in Appendix \ref{fullefficacy}, \ref{fullphasic} and \ref{fullnum} respectively.}
    \label{fig:normalized-returns}
\end{figure}

\paragraph{Efficacy of MAPO-LSO} As depicted in Figure \ref{normalized-efficacy}, the MAPO-LSO framework demonstrates a significant improvement in the collective return, reaching a $\mathbf{+35.68\%}$ difference in convergence from the baseline without MA-LSO. Additionally, in terms of sample efficiency, our MAPO-LSO achieved the max convergence of the baseline in $\mathbf{285.7\%}$ less samples. However, to achieve this, we discuss the design choices made that enabled this improvement.

\subsubsection{Design Choices in MAPO-LSO} 

\paragraph{Phasic Optimization} We confirm our hypothesis stated in Section \ref{phasicdescr} with Figure \ref{normalized-phasic}, where we found training inefficiencies with a shared encoder between the actors and critics in the MARL algorithms (i.e. HAPPO, MADDPG and MASAC) with phasic learning. Aforementioned, this concern is not novel \cite{cobbe2021phasic} and in our work, we addressed this issue through phasic regularization with significant improvements. Moreover, we encourage future works to explore further methodologies that can facilitate a shared encoder paradigm, as we did find that the robustness of hyperparameters can still be improved upon.

\begin{wrapfigure}{r}{0.4\textwidth}
    \centering
    \includegraphics[width=0.4\textwidth]{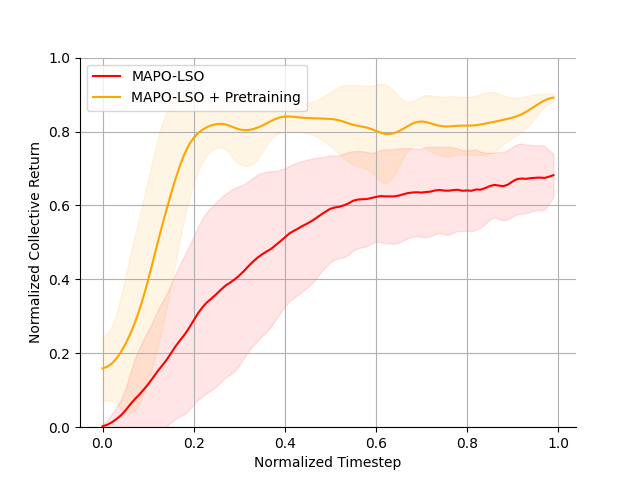}
    \caption{MAPO-LSO as a pre-training process is evaluated, normalized on all runs listed in Appendix \ref{fullpt} with the error bars showing the $\pm1$ std deviation.}
    \label{fig:pretraining}
\end{wrapfigure}

\begin{figure}
    \centering
    \subfloat[Belief Space\label{normalized-beliefs}]{\includegraphics[width=0.5\textwidth]{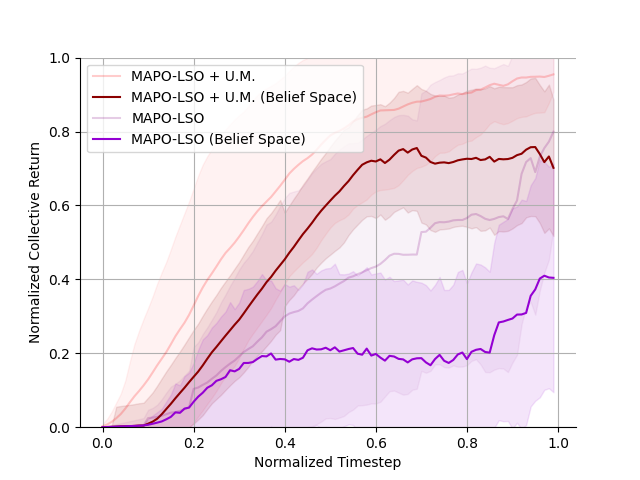}}
    \subfloat[MA-TDR Losses\label{normalized-tdr}]{\includegraphics[width=0.375\textwidth]{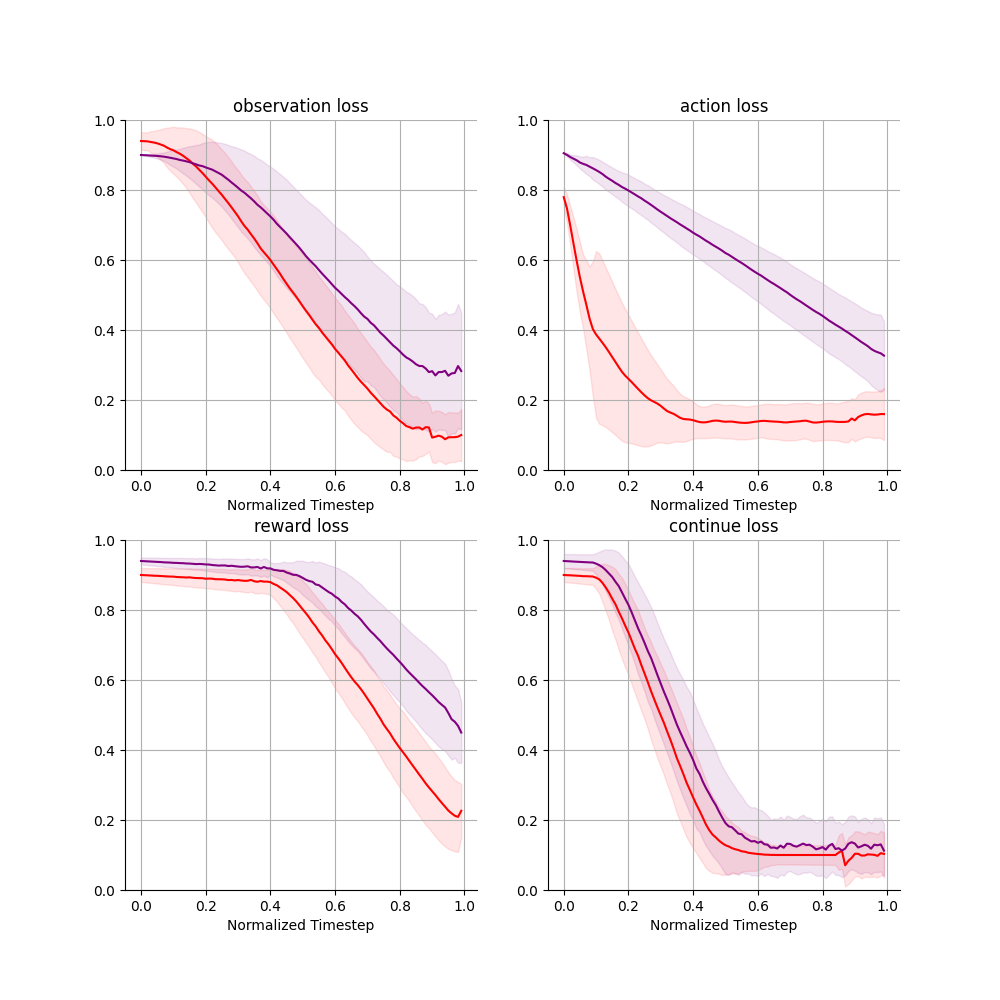}}
    \caption{\textbf{Left:} A comparison of normalized collective returns between MAPO-LSO with and without uncertainty modeling (U.M.), where the actions inferred by each agent's belief space are used. We normalize the sum of the collective returns over all agent's imagined joint policies on the VMAS and IST benchmarks. \textbf{Right:} The four graphs show the normalized MA-TDR losses between MAPO-LSO using and not using uncertainty modeling. For both plots, the error bars indicate $\pm1$ std deviations over all runs.}
    \label{fig:uncertainty}
\end{figure}

\paragraph{Epistemic Uncertainty Modeling} Referring to Figure \ref{normalized-tdr}, the uncertainty modeling within the MA-TDR heads demonstrates improvements in the accuracy of the beliefs of observation, action, reward and continue signals, most notably having the largest impact on the accuracy of inferring the actions. Furthermore, we evaluate the imagined policies realized within each agent's belief space by rolling out trajectories using the joint actions within the agent's belief space. We find the uncertainty modeling does influence the behaviors learned within each agent's belief spaces, as shown in Figure \ref{normalized-beliefs}, exhibiting impressive performance even using these imagined joint policies. Unsurprisingly, this uncertainty modeling also improved the expected collective return as well, as seen in Figure \ref{normalized-num}. In future works, a further evaluation and study into the diversity and social behaviors learned within these imagined joint policies would be fruitful.

\paragraph{Pre-training} We study the efforts of using MA-TDR and MA-SPL objectives as a pre-training process. First, we collected a dataset of $10$K trajectories using a random policy and pre-trained the model on the MA-LSO objectives for $100$ epochs. As shown in Figure \ref{fig:pretraining}, we find that the inclusion of pre-training provides an improvement of $+21.0\%$ in the collective return achieved.

\begin{table}[!htbp]
\centering
\begin{tabular}{c|cccc|c}\\ 
       & $\mathscr{L}_{tdr}$ & \multicolumn{3}{c}{$\mathscr{L}_{spl}$} &                    \\
case      &        & $\mathscr{L}_{\text{MA-MR}}$  & $\mathscr{L}_{\text{MA-FDM}}$  & $\mathscr{L}_{\text{MA-IDM}}$  & Max Return \\ \hline \hline
MA-LSO & $0.023 \pm 0.055$ & $0.033 \pm 0.028$ & $0.058 \pm 0.025$ & $0.048 \pm 0.022$ &
 $0.954 \pm 0.046$ \\
no MA-TDR    & $0.895 \pm 0.079$ & $0.288 \pm 0.104$ & $0.192 \pm 0.138$ & $0.191 \pm 0.159$ &  $0.847 \pm 0.112$ \\
no MA-MR  & $0.188 \pm 0.129$ & $0.716 \pm 0.086$ & $0.164 \pm 0.124$ & $0.102 \pm 0.131$ & $0.887 \pm 0.122$ \\
no MA-FDM & $0.092 \pm 0.110$ & $0.165 \pm 0.030$ & $0.726 \pm 0.073$ & $0.159 \pm 0.128$ &      $0.911\pm 0.185$ \\
no MA-IDM & $0.152 \pm 0.134$ & $0.247 \pm 0.055$ & $0.193 \pm 0.104$ & $0.793 \pm 0.124$ &   $0.904\pm 0.114$   \\ 
no MA-SPL & $0.331 \pm 0.127$ & $0.848 \pm 0.132$ & $0.899 \pm 0.053$ & $0.933 \pm 0.092$ &   $0.819\pm 0.198$   \\ 
no MA-LSO & $0.970 \pm 0.054$ & $0.964 \pm 0.088$ & $0.958 \pm 0.080$ & $0.908 \pm 0.072$ &  $0.598\pm0.201$   \\ \hline\hline
\end{tabular}
\caption{Empirical results from our ablation studies on the components of MA-LSO, comparing the loss terms and the maximum normalized return achieved with its respective $\pm 1$ std deviation. For more details, refer to Appendix \ref{fullabalation}.}\label{table}
\end{table}

\paragraph{MA-LSO Ablations} We assess the effectiveness of each component within our MA-LSO framework by conducting evaluations that include omissions of the MA-TDR and MA-SPL processes. For MA-SPL, we exclude its sub-processes individually: MA-MR, MA-FDM, and MA-IDM. A key contribution of this work is the integration of these auxiliary objectives and their symbiotic relationship, which Table \ref{table} confirms. Hence, the results demonstrate that all of the components in our MA-LSO framework not only contribute to the demonstrated improvements but also are interdependent. Specifically, MA-SPL has the greatest impact in terms of overall performance, with MA-MR being the most important out of its sub-processes. This highlights the importance of the relational information instilled by MA-SPL and MA-MR and the consistency they endow within the latent state space between the agents.

Moreover, excluding any processes within MA-LSO results in a notable decline in the training efficiency of other processes. This suggests a form of amortization similar to that observed in multi-task applications \cite{kalashnikov2021mt}, evident from optimal performance of each component is only achieved when both MA-TDR and MA-SPL are applied in unison. The interdependence of these components is underscored by the fact that the convergence of MA-TDR and MA-SPL losses deteriorates when they are separated. Specifically, without MA-SPL, the convergence of MA-TDR decreases by $30.9\%$, while the absence of MA-TDR leads to degradation of MA-SPL subprocesses by $25.5\%$, $13.4\%$, and $14.3\%$ on MA-MR, MA-FDM, and MA-IDM respectively.




\section{Conclusion}
We introduce a generalized MARL training paradigm, MAPO-LSO, that utilizes auxiliary learning objectives to enrich the MARL learning process with multi-agent transition-dynamics reconstruction and self-predictive learning. Our approach improves its "non-LSO" counterpart in a wide variety of MARL benchmark tasks using several state-of-the-art MARL algorithms. For future directions, there remain promising avenues to study other aspects of the multi-agent nature of MARL tasks, such as ad-hoc performance and social learning, with our MAPO-LSO framework.

\bibliography{neurips_2024}

\newpage
\appendix

\section{MARL Environments}\label{alltasks}
\subsection{Vectorized Multi-Agent Environments}
\begin{figure}[!htb]
    \centering
    \includegraphics[width=0.7\textwidth]{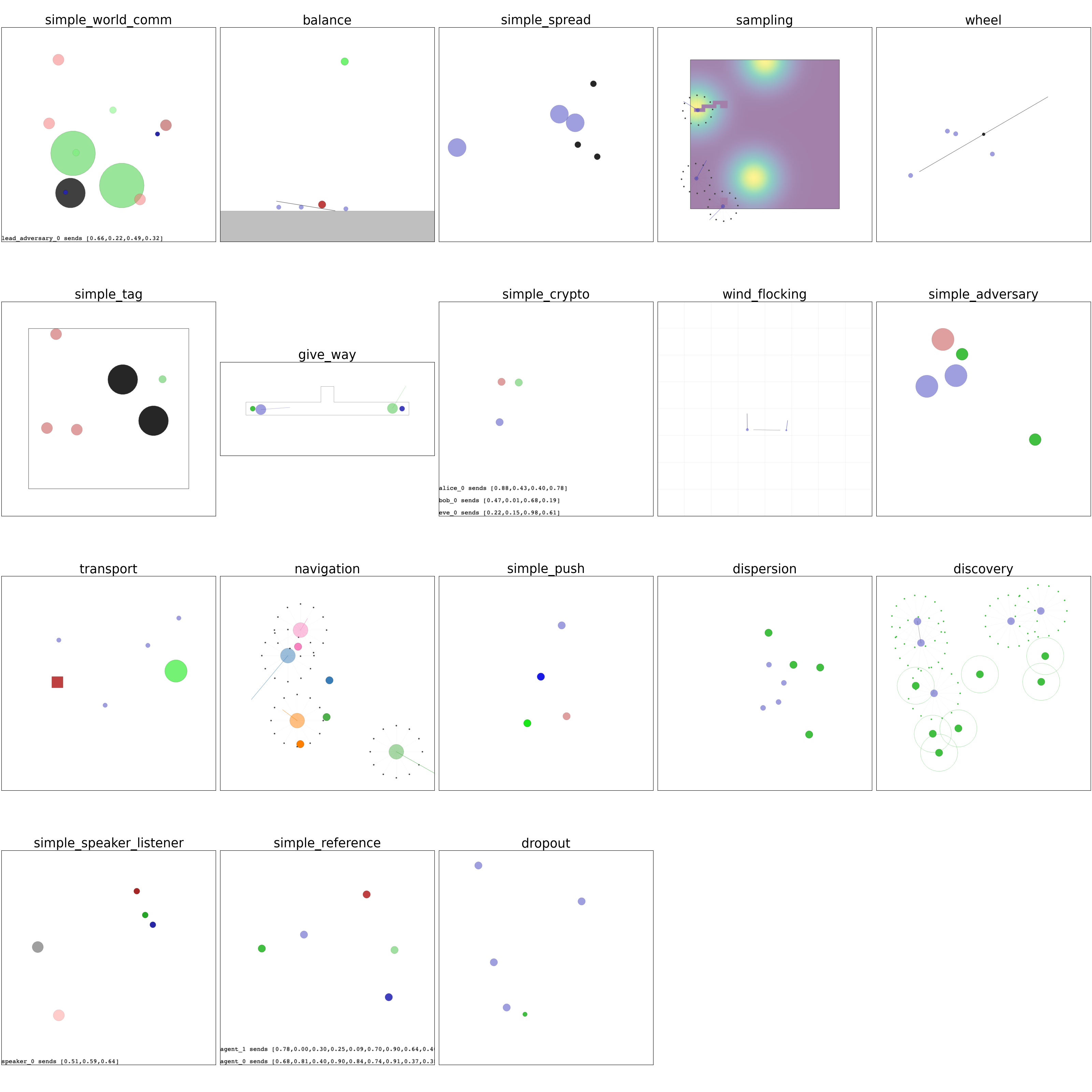}
    \caption{The 18 VMAS tasks used for our evaluations. Their full descriptions can be found in \cite{bettini2022vmas}. For each task, we use $1024$ parallel environments for training and $16$ for evaluation and ran the training for $[100t, 200t, 500t, 1000t]$ time-steps, depending on the learning performance, where $t$ is the time horizon for each task.}
    \label{fig:enter-label}
\end{figure}
\newpage
\subsection{IsaacTeams}
\begin{figure}[!htb]
    \centering
    \subfloat[abb-reacher-2]{\includegraphics[width=0.3\textwidth] {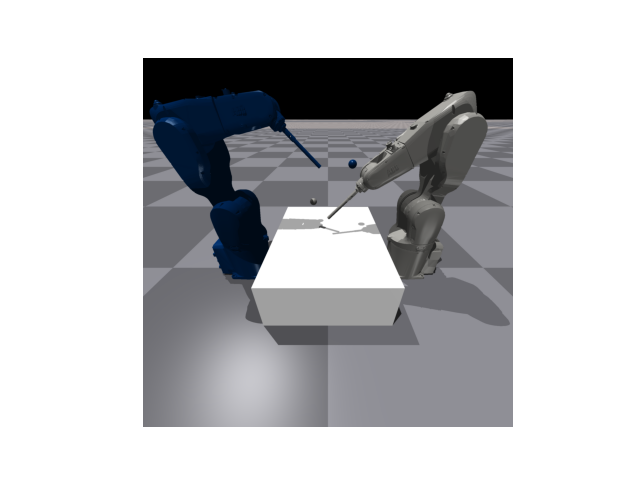}} \quad
    \subfloat[franka-reacher-2]{\includegraphics[width=0.3\textwidth]{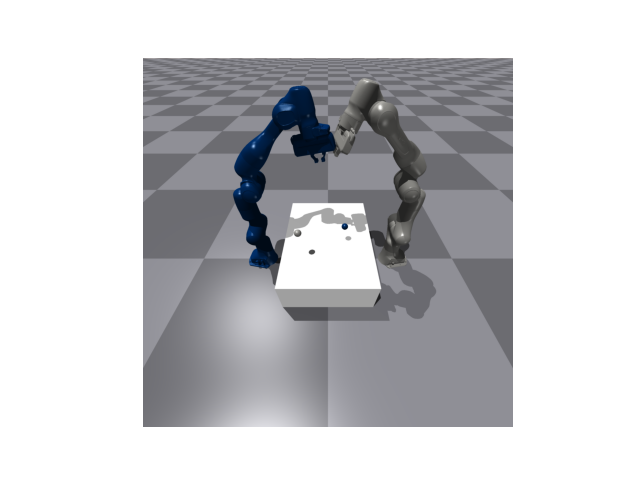}}\quad
    \subfloat[kuka-reacher-2]{\includegraphics[width=0.3\textwidth]{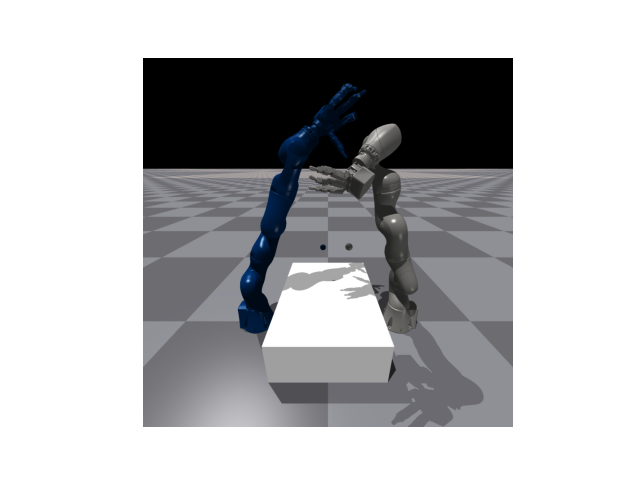}}\quad
    \subfloat[afk-reacher-3]{\includegraphics[width=0.3\textwidth]{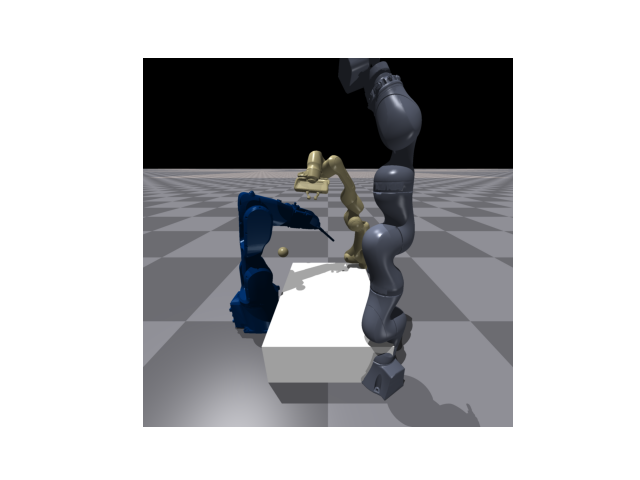}}\quad
    \subfloat[visual-afk-reacher-3\label{visualreacher}]{\includegraphics[width=0.3\textwidth]{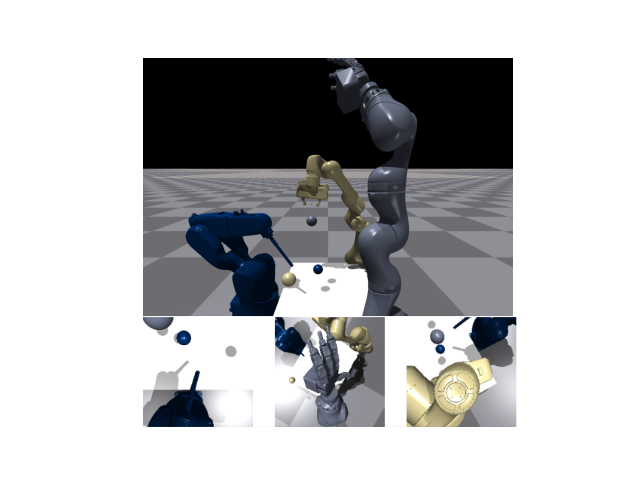}}
    \caption{The 5 IST tasks used for our evaluation. For all tasks, the objective is to move the end-effectors to their respective target spheres, where the reward function is shaped to minimize $\ell_2$ distance. These target spheres are positioned randomly. The observation space include the robotic arm's proprioceptive information as well as the information regarding its target sphere. For the visual-afk-reacher-3 task, the visual input (with resolution of $32\times32$) that is shown at the bottom of Figure \ref{visualreacher} is appended to the observation space of the respective agent, where the visualization shown in Figure \ref{visualreacher} has increased resolution for presentation purposes. The action space controls the joint actuation of the robotic arms. All tasks define a communication network that enables full communication. The training procedure otherwise follows the one set for VMAS, except for the visual-afk-reacher-3, where we use $128$ parallel environments for training.}
    \label{fig:enter-label}
\end{figure}

\newpage
\section{Implementation of MARL Algorithms}
\paragraph{MAPPO} Multi-Agent Proximal Policy Optimization (MAPPO) \cite{yu2022surprising} is a CTDE extension of the Proximal Policy Optimization (PPO) algorithm that employs decentralized policies with centralized value functions. In our implementation, we follow the original paper's implementation but with a centralized critic shared between all agents.

\paragraph{HAPPO} Heterogenous-Agent Proximal Policy Optimization \cite{kuba2021trust} refines MAPPO, imposing a random-order sequential-update scheme to ensure monotonic improvements unrestricted to the assumption of homogeneity of agents. Our implementation follows the original work, but the main differences stem largely from the shared encoder between the policy and value function. To ensure more stable learning, largely due to the shared encoder, we update the value function upon each agent update and reduce the learning rate of the encoder. 

\paragraph{MADDPG} Similar to MAPPO, Multi-Agent Deep Deterministic Policy Gradient (MADDPG) \cite{lowe2017multi} is a CTDE extension of the Deep Deterministic Policy Gradient (DDPG) algorithm. The main difference in our implementation follows \cite{fujimoto2018addressing}, including delayed policy updates, target policy smoothing, clipped double learning, stochastic actors and a shared critic between all agents.

\paragraph{MASAC} Multi-Agent Soft Actor Critic is a CTDE extension of the Soft Actor Critic (SAC) algorithm \cite{haarnoja2018soft}. Our implementation is similar to our MADDPG implementation with adjustments for auto-tuned entropy maximization.

\section{Model Architecture}\label{appendix:model}
For all algorithms, we follow the same model architecture we described below.

\paragraph{Encoder} The encoder is responsible for embedding the input data and follows the DreamerV3 encoder architecture \cite{hafner2023mastering} that most matches the 12M parameter model. For multi-modal data, we process the different modality of data separately, i.e. images with a CNN and structured data with a MLP, and aggregate the embeddings with a sum operator. We define a separate encoder for each agent.

\paragraph{Communication Block} The communication block propagates the embeddings between agents dependent on the communication graph to produce the latent space. We modeled this component after MAMBA's communication block \cite{egorov2022scalable}, although we opted to have a smaller model. For partial communication graphs, we mask the embeddings of the unconnected agents. The policy of each agent uses their own latent state to compute their actions, and the centralized critic concatenates the latent state of all agents to compute the value for all agents.

\paragraph{MA-TDR} Similar to the encoder, the components such as the decoder and the action/reward/continue (ARC) heads were all modeled following the DreamerV3 architecture \cite{hafner2023mastering}. The ARC heads that modeled beliefs of other agents were appended with an MC-Dropout layer \cite{gal2016dropout} when uncertainty modeling is used.

\paragraph{MA-SPL} For MA-MR, we follow the same setup as CURL \cite{laskin2020curl} and for MA-FDM and MA-IDM, we mostly adhere to the same procedure and model architecture as single-agent SPL \cite{schwarzer2021pretraining} with random noise augmentation. For $\mathcal{R}, \mathcal{T}_z, \mathcal{I}$, we use a multi-headed attention head, where:
$$\mathcal{R}(m_i(z_t)) = \text{MultiHeadedAttn}(q = m_i(z_t), k=m_i(z_t), v= m_i(z_t))$$
$$\mathcal{T}_z(z_t, a_t) = \text{MultiHeadedAttn}(q = a_t, k=z_t, v= z_t)$$
$$\mathcal{I}(z_t, z_{t+1}) = \text{MultiHeadedAttn}(q = \llbracket z_t, z_{t+1} \rrbracket, k=\llbracket z_t, z_{t+1} \rrbracket, v= \llbracket z_t, z_{t+1} \rrbracket)$$
where $\llbracket z_t, z_{t+1} \rrbracket$ concatenates $z_t$ and $ z_{t+1}$ into a single sequence.
\newpage
\section{Hyperparameters}\label{hyperparameters}
For each task, we initially ran random search over the following hyperparameters and followed up with further tuning using qualitative examinations over these runs.

\begin{table}[!htb]
    \begin{minipage}{\textwidth}
      \caption{MAPPO/HAPPO}
      \centering
        \begin{tabular}{p{0.5\textwidth} | p{0.5\textwidth}}
          Name  & Value \\ \hline
          learning rate & [1e-3, 5e-4, 1e-4, 5e-5, 1e-5, 1e-6] \\
          entropy coef. & [1e-5, 1e-3, 3e-4] \\
          clip coef. & [0.05, 0.1, 0.15, 0.2, 0.3, 0.5] \\
          discount factor & 0.99 \\
          num. of updates & 30 \\
          target KL & [0.01, None] \\
          gradient norm & [0.5, 1.0, None] \\
          $\lambda$-return & 0.95
        \end{tabular}
    \end{minipage}%

    \begin{minipage}{\linewidth}
     \caption{MADDPG}
      \centering
        \begin{tabular}{p{0.5\textwidth} | p{0.5\textwidth}}
          Name  & Value \\ \hline
          learning rate & [1e-3, 5e-4, 1e-4, 5e-5, 1e-5, 1e-6] \\
          exploration noise & [0.01, 0.1, 0.5] \\
          learning starts & [t, 2t] \\
          smoothing noise & [0.1, 0.2, 0.5]\\
          smoothing noise clip & [0.01, 0.1] \\
          num. of updates & 50 \\
          policy update frequency & 2\\
          gradient norm & [0.5, 1.0, None] \\
          target network update $\tau$ & 0.005 \\
          target network frequency & 2
        \end{tabular}
    \end{minipage} 

    \begin{minipage}{\linewidth}
     \caption{MASAC}
      \centering
        \begin{tabular}{p{0.5\textwidth} | p{0.5\textwidth}}
          Name  & Value \\ \hline
          learning rate & [1e-3, 5e-4, 1e-4, 5e-5, 1e-5, 1e-6] \\
          learning starts & [t, 2t] \\
          num. of updates & 50 \\
          policy update frequency & 2\\
          gradient norm & [0.5, 1.0, None] \\
          target network update $\tau$ & 0.005  \\
          target network frequency & 2
        \end{tabular}
    \end{minipage} 

    \begin{minipage}{\linewidth}
     \caption{MA-TDR (For MAPO-LSO)}
      \centering
        \begin{tabular}{p{0.5\textwidth} | p{0.5\textwidth}}
          Name  & Value \\ \hline
          $\alpha_{\text{obs}}$ & [1e-3,0.1,0.5,1] \\
          $\alpha_{\text{act}}$ & [1e-3,0.1,0.5,1] \\
          $\alpha_{\text{rew}}$ & [1e-3,0.1,0.5,1] \\
          $\alpha_{\text{cont}}$ & [1e-3,0.1,0.5,1] \\
          $\alpha_{\text{curl}}$ & [1e-3,0.1,0.5,1] \\
          dropout & [0, 0.1, 0.2, 0.5, 0.8]
        \end{tabular}
    \end{minipage} 
    
    \begin{minipage}{\linewidth}
     \caption{MA-SPL (For MAPO-LSO)}
      \centering
        \begin{tabular}{p{0.5\textwidth} | p{0.5\textwidth}}
          Name  & Value \\ \hline
          $\alpha_{\text{mr}}$ & [1e-3,0.1,0.5,1] \\
          $\alpha_{\text{fdm}}$ & [1e-3,0.1,0.5,1] \\
          $\alpha_{\text{idm}}$ & [1e-3,0.1,0.5,1]
        \end{tabular}
    \end{minipage} 
    \caption{Hyperparameters for MAPPO, HAPPO, MADDPG, and MASAC, and the MA-LSO learning processes, where $t$ is the length of a full trajectory. The batch-size was set based on the maximum load possible on our GPU, which differed for all tasks. For MAPO-LSO trainings, the hyperparameters for the MARL algorithms are fixed.}
\end{table}

\newpage
\section{Full Results: Efficiacy of MAPO-LSO} \label{fullefficacy}
\begin{figure}[!htbp]
    \centering
    \subfloat{{\includegraphics[width=0.45\textwidth]{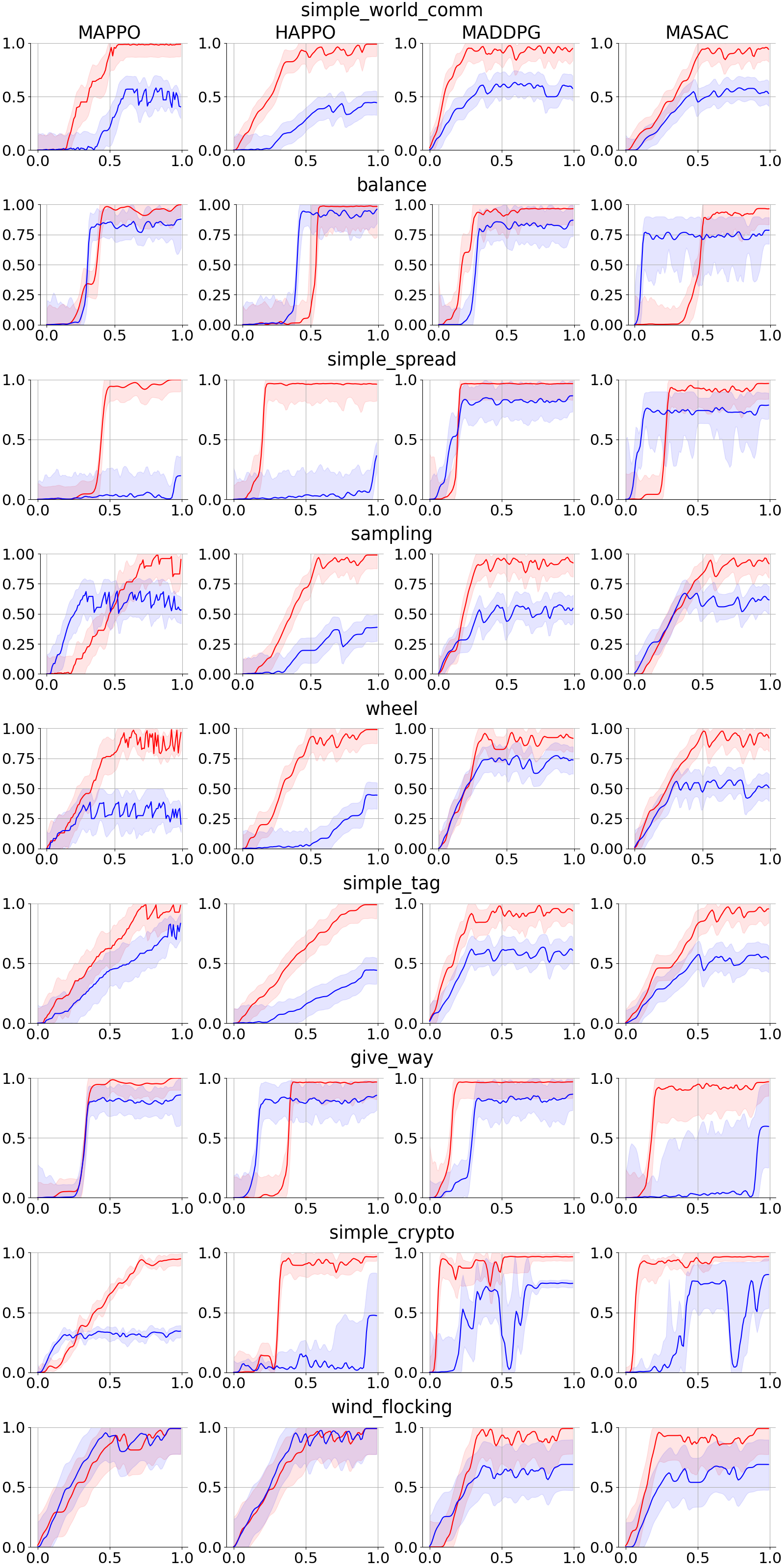} }}%
    \qquad
    \subfloat{{\includegraphics[width=0.45\textwidth]{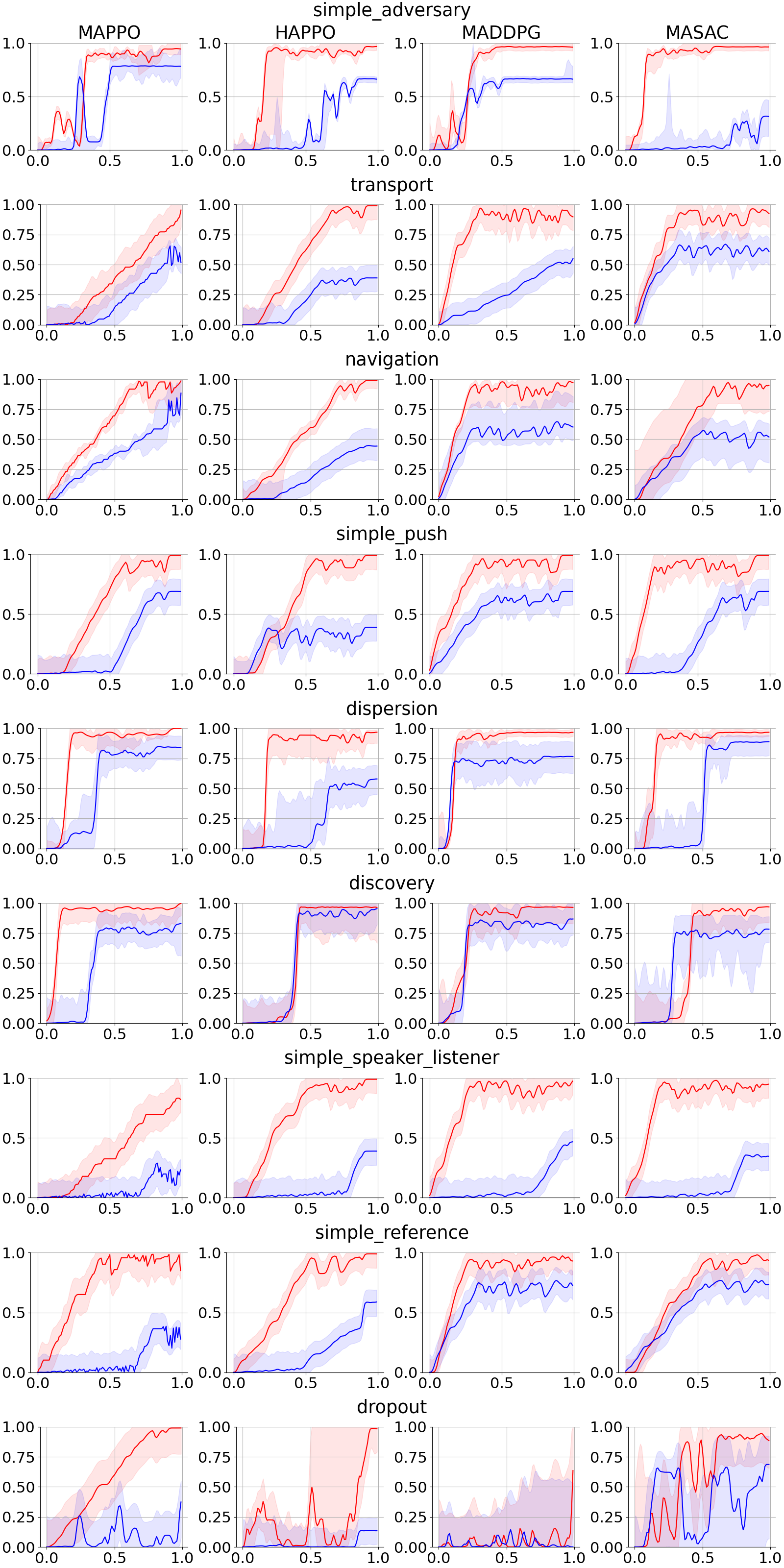} }}%
    \qquad
    \subfloat{{\includegraphics[width=0.45\textwidth]{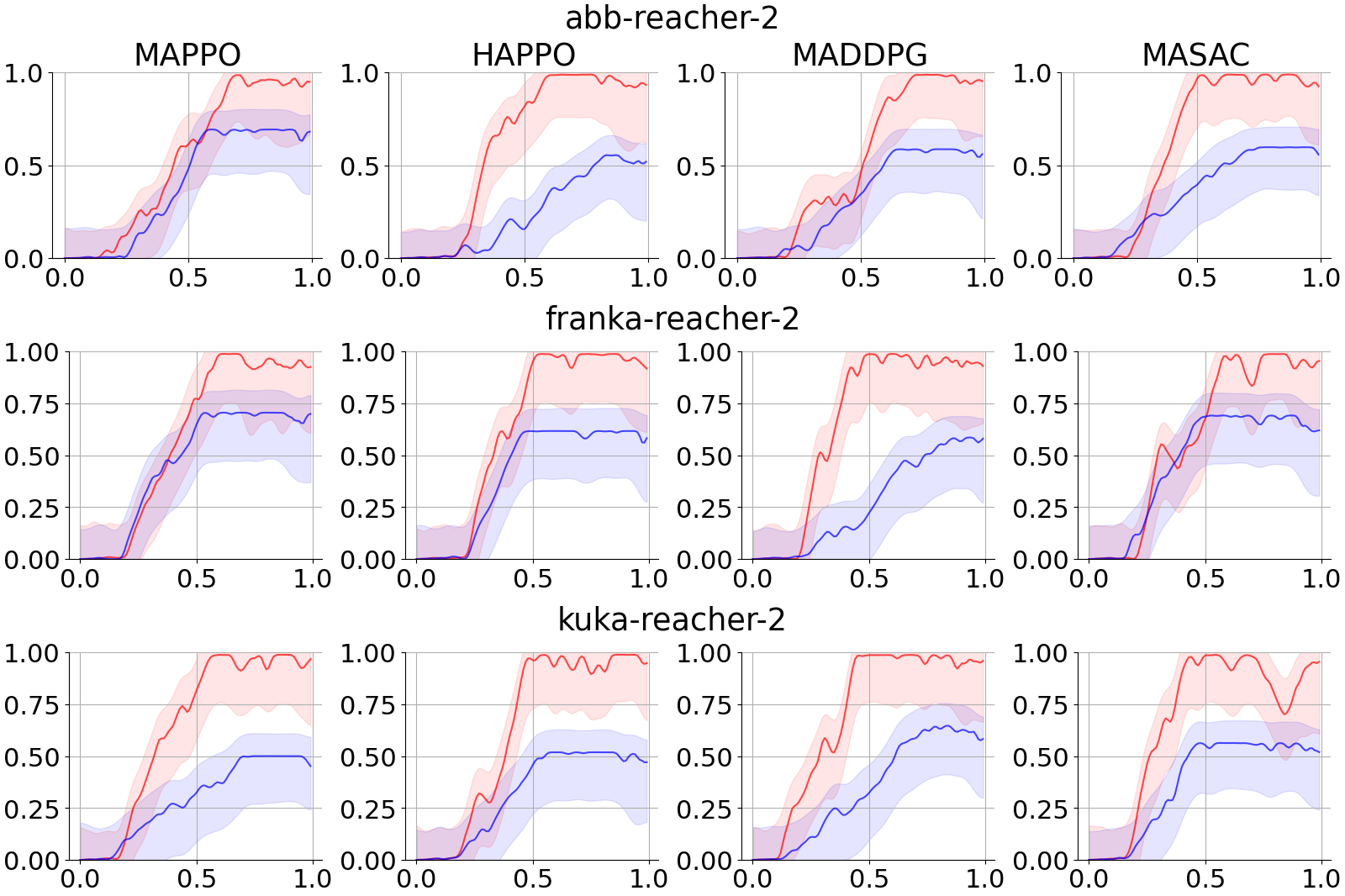} }}%
    \qquad
    \raisebox{4em}{\subfloat{{\includegraphics[width=0.45\textwidth]{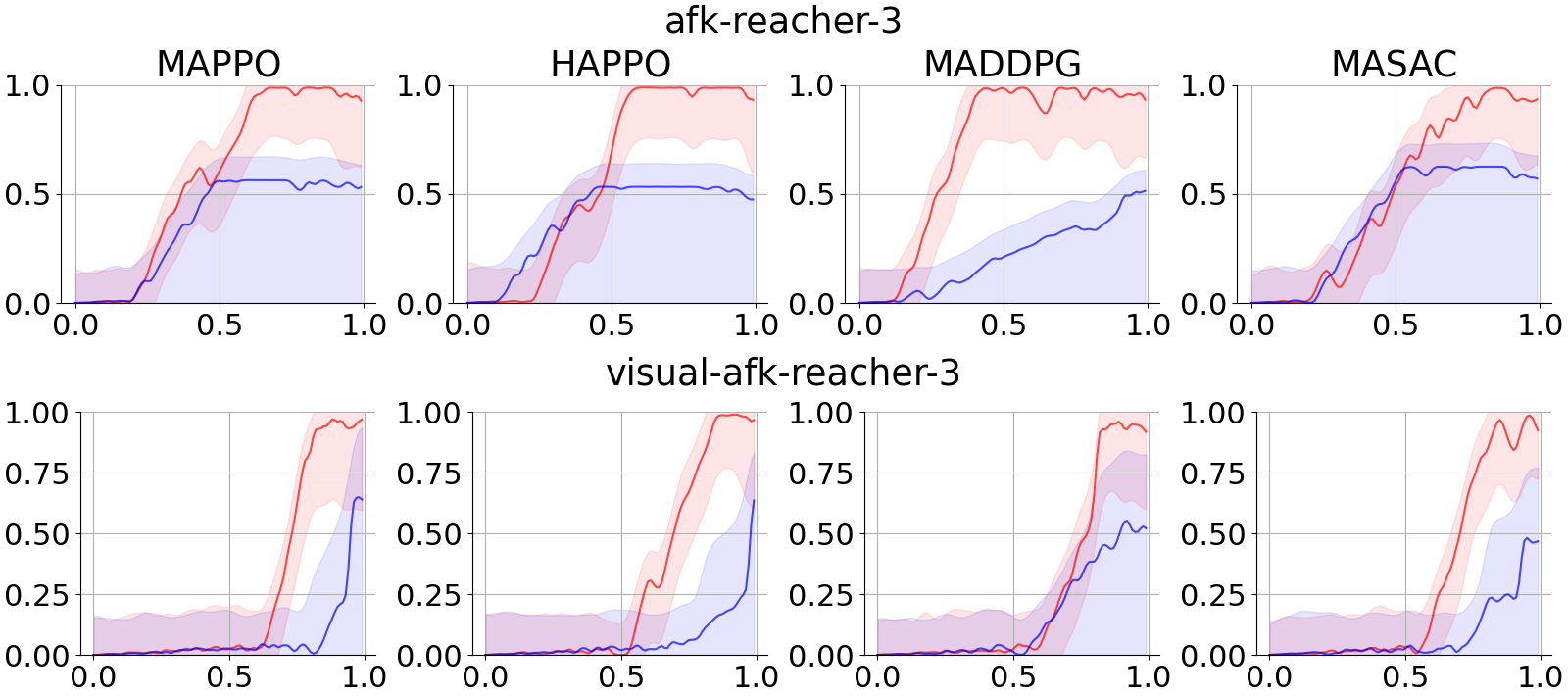} }}}%
    \caption{VMAS/IST Results For MAPO-LSO: These plots compare the performance of the traditional MARL algorithms (shown in the blue line) versus its LSO counterpart (shown in the red line) in each task tested in the VMAS/IST benchmark under a normalized scale. Each column of plots uses the same MARL algorithm and each row evaluates on the same task. The y-axis is the normalized collective return and the x-axis is the normalized time-steps, with evaluation ran over $16$ random seeds. The error bars show the min and max returns over those $16$ runs.}%
    \label{fig:fullist}%
\end{figure}

\newpage
\section{Full Results: Phasic Optimization For HAPPO/MADDPG/MASAC}  \label{fullphasic}
\begin{figure}[!htbp]
    \centering
    \subfloat{{\includegraphics[width=0.45\textwidth]{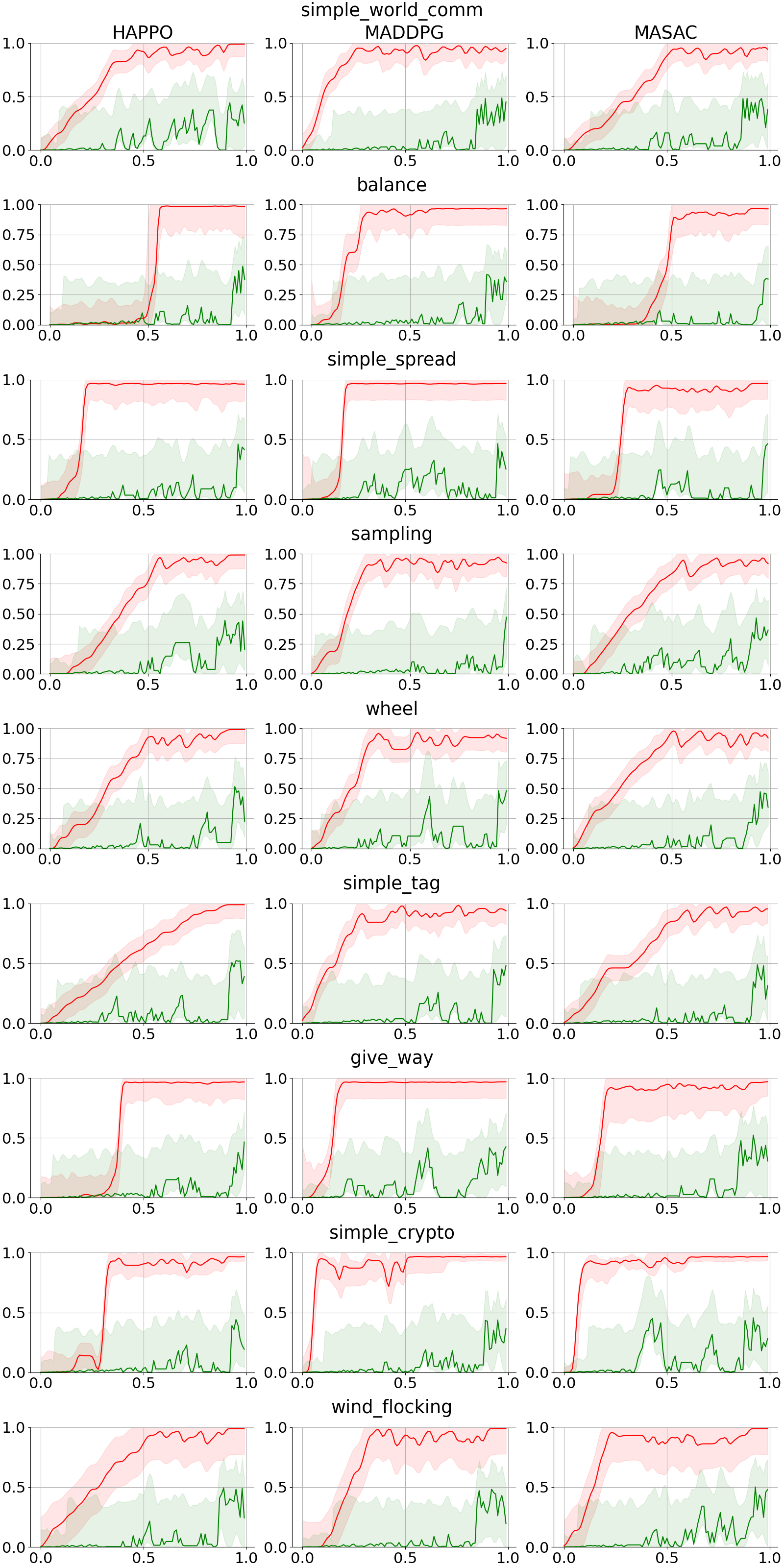} }}%
    \qquad
    \subfloat{{\includegraphics[width=0.45\textwidth]{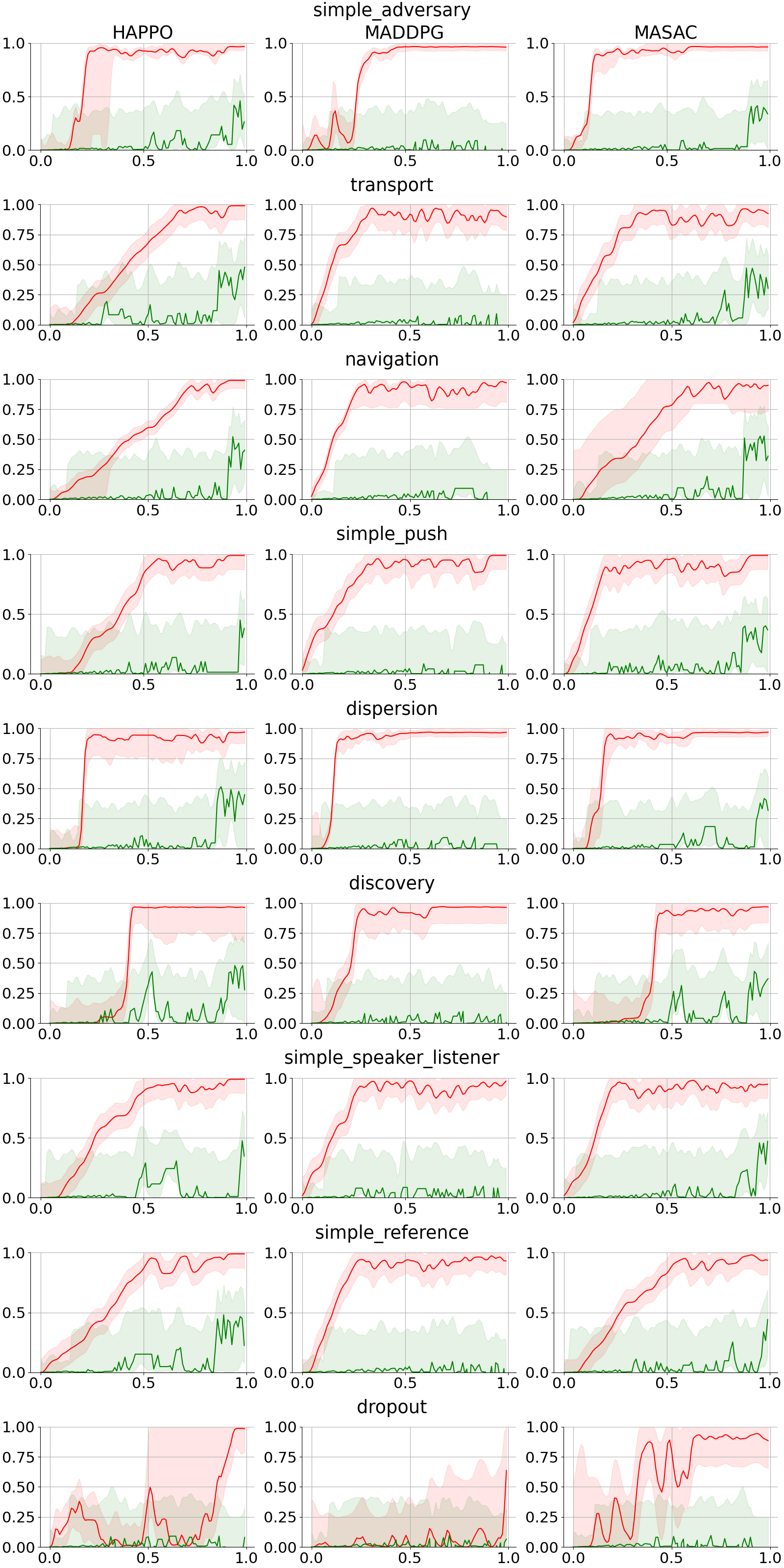} }}%
    \qquad
    \subfloat{{\includegraphics[width=0.45\textwidth]{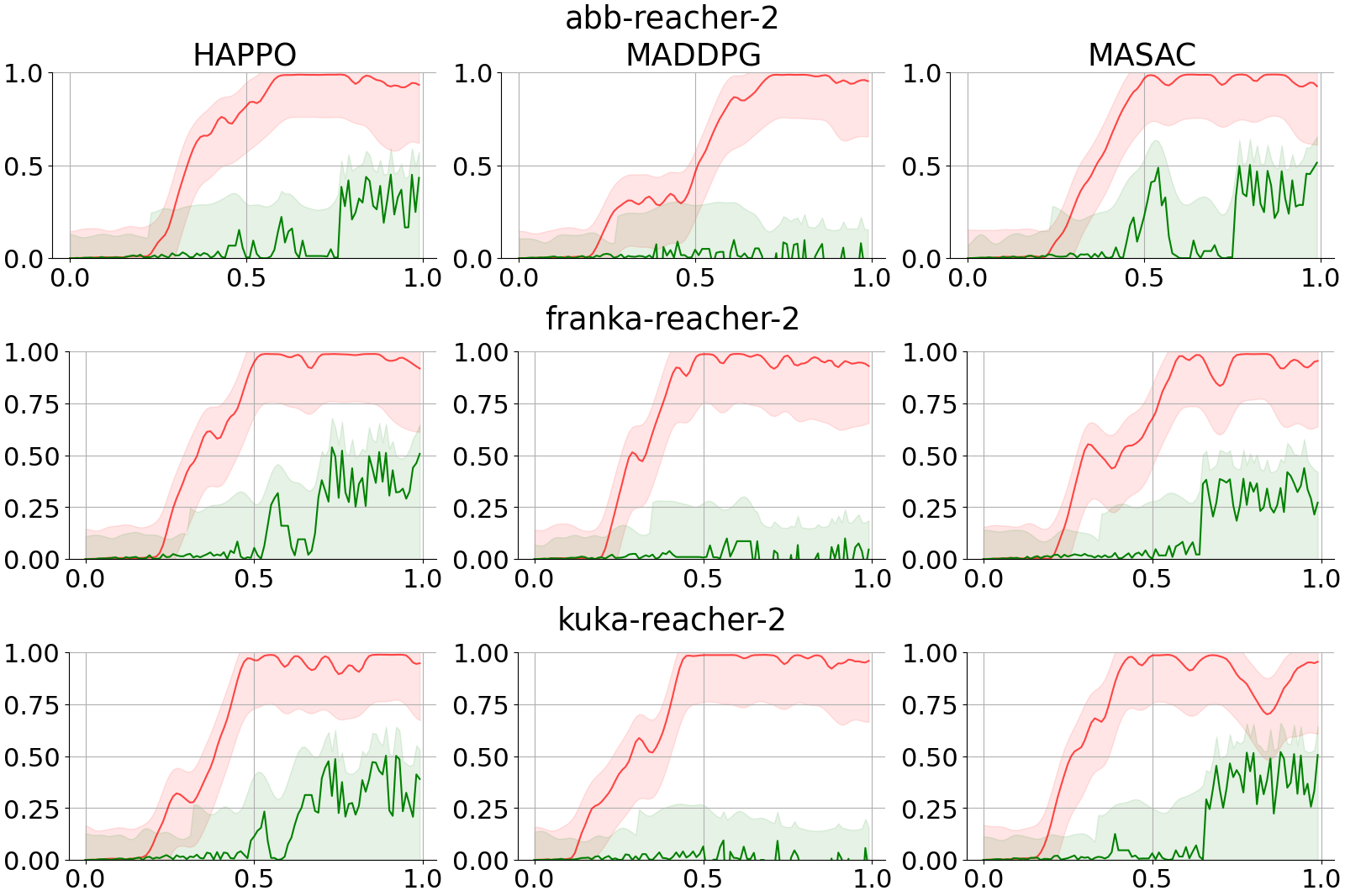} }}%
    \qquad
    \raisebox{4em}{\subfloat{{\includegraphics[width=0.45\textwidth]{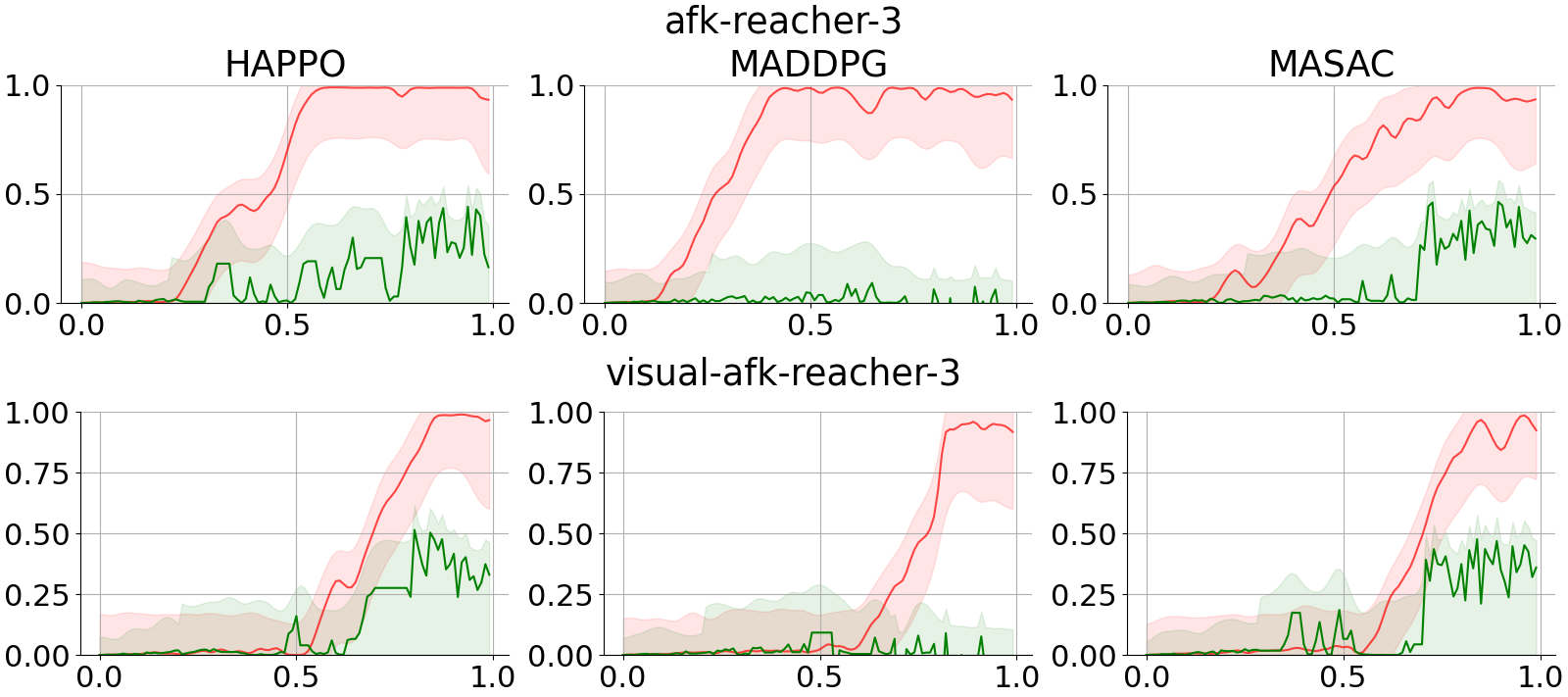} }}}%
    \caption{VMAS/IST Results For Phasic Optimization: These plots compares the performance of MAPO-LSO with (shown in the red line) and without (shown in the green line) phasic optimization in each task tested in the VMAS/IST benchmark under a normalized scale. We note that the same hyperparameters are used for both, but they are tuned for MAPO-LSO \textbf{with} phasic optimization. The format follows Figure \ref{fig:fullist}.}%
    \label{fig:fullphasic}%
\end{figure}

\newpage
\section{Full Results: Pretraining Experiments} \label{fullpt}
\begin{figure}[!htbp]
    \centering
    \subfloat{{\includegraphics[width=0.45\textwidth]{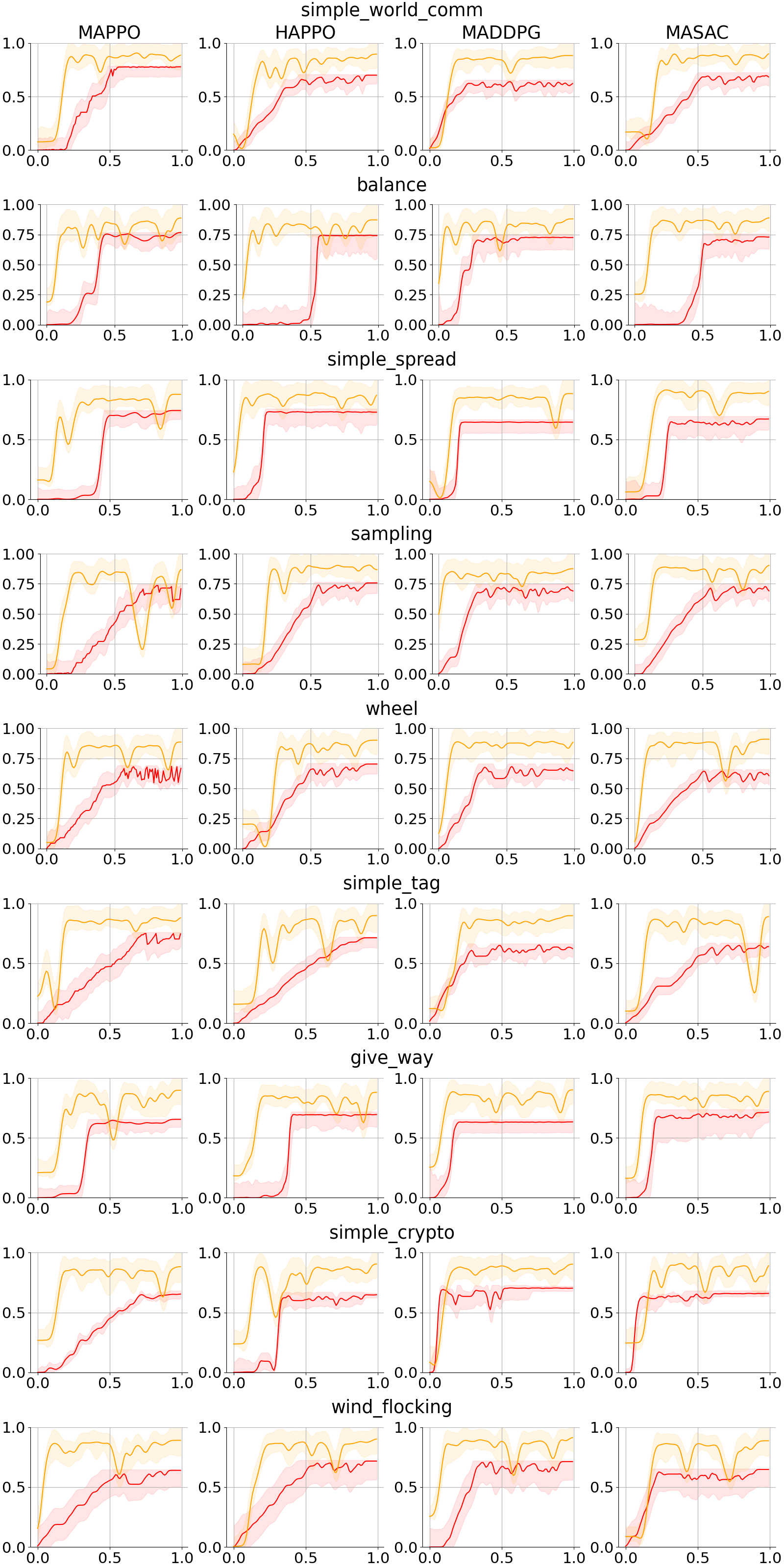} }}%
    \qquad
    \subfloat{{\includegraphics[width=0.45\textwidth]{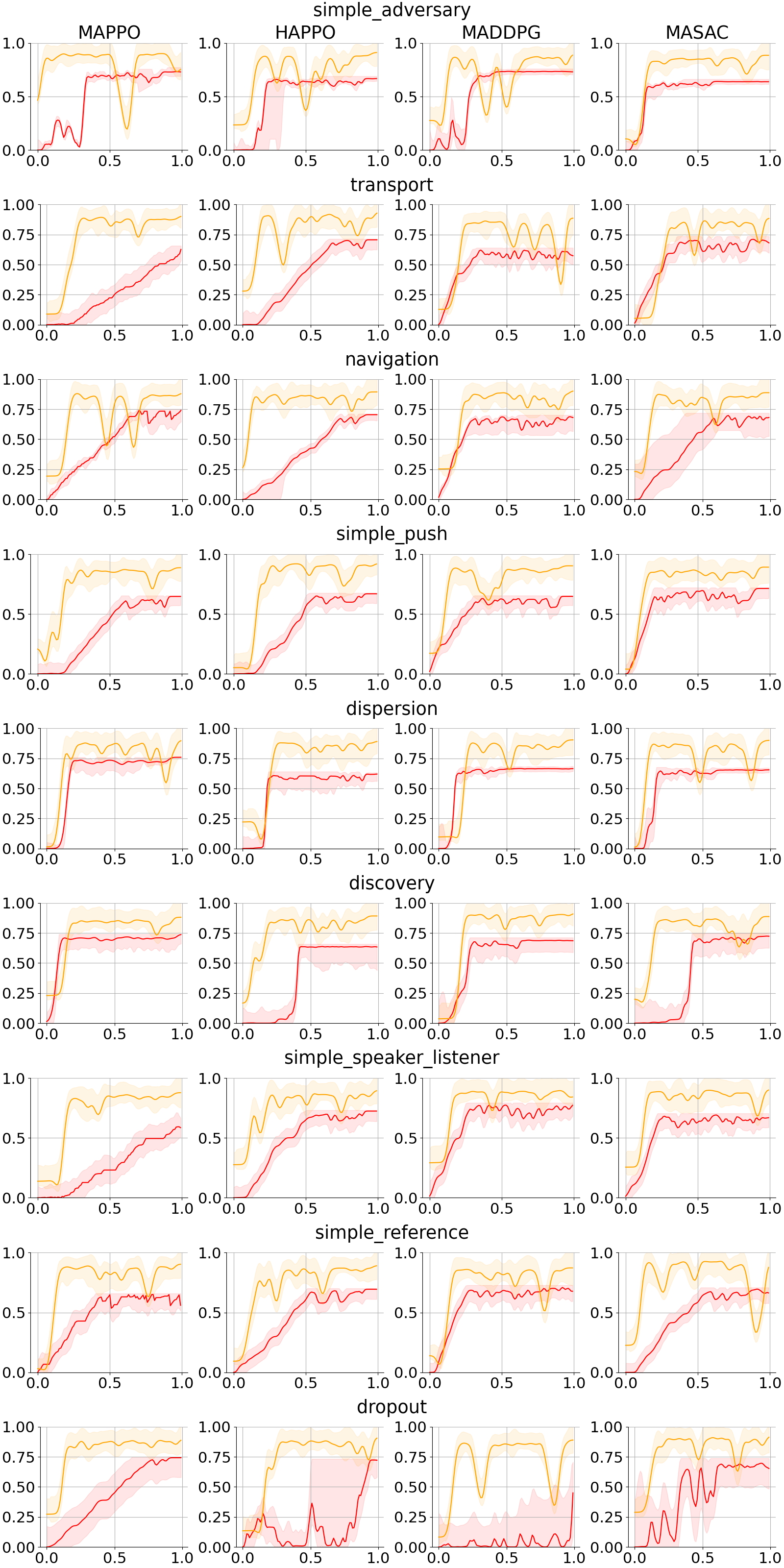} }}%
    \qquad
    \subfloat{{\includegraphics[width=0.45\textwidth]{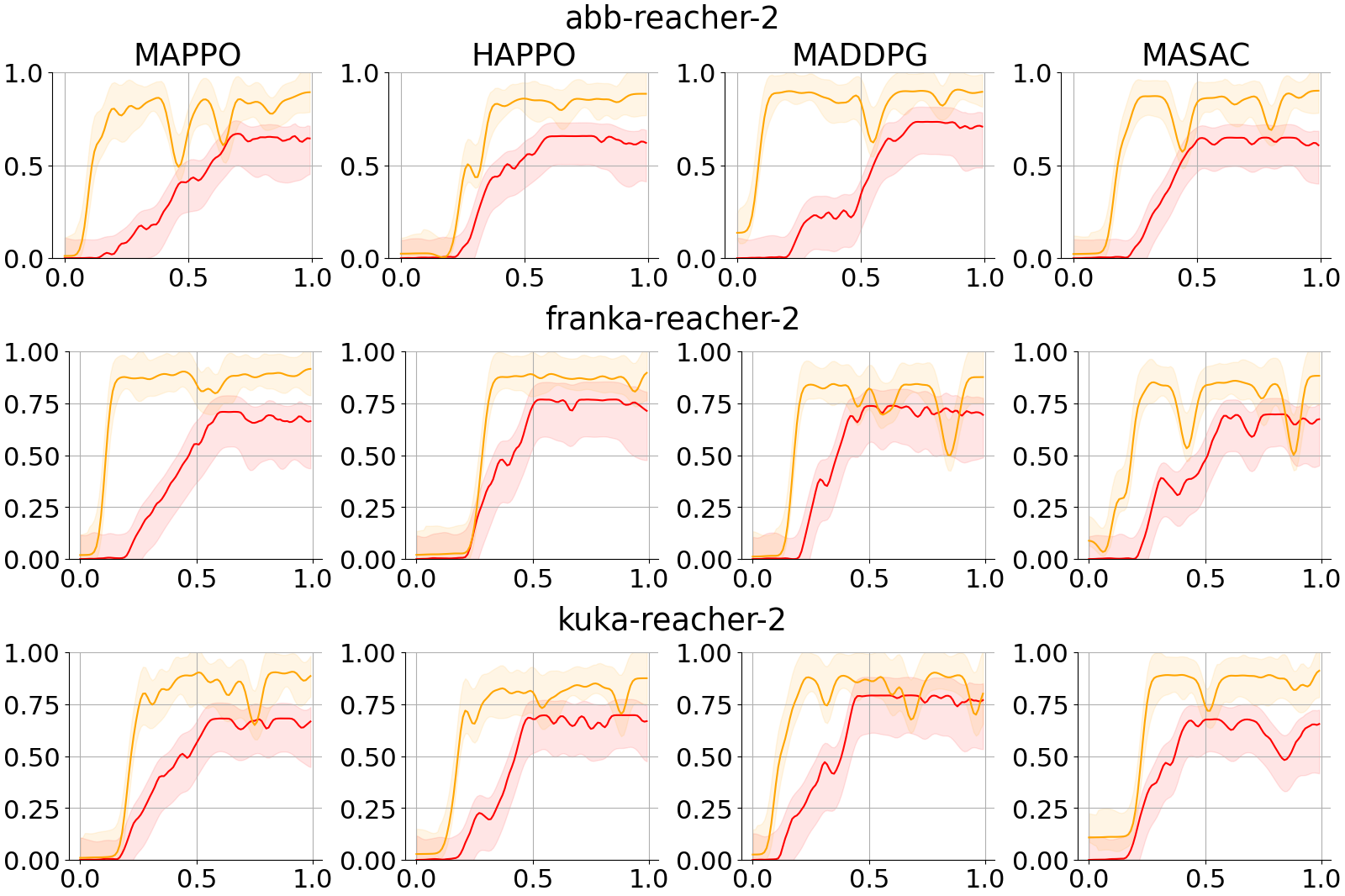} }}%
    \qquad
    \raisebox{4em}{\subfloat{{\includegraphics[width=0.45\textwidth]{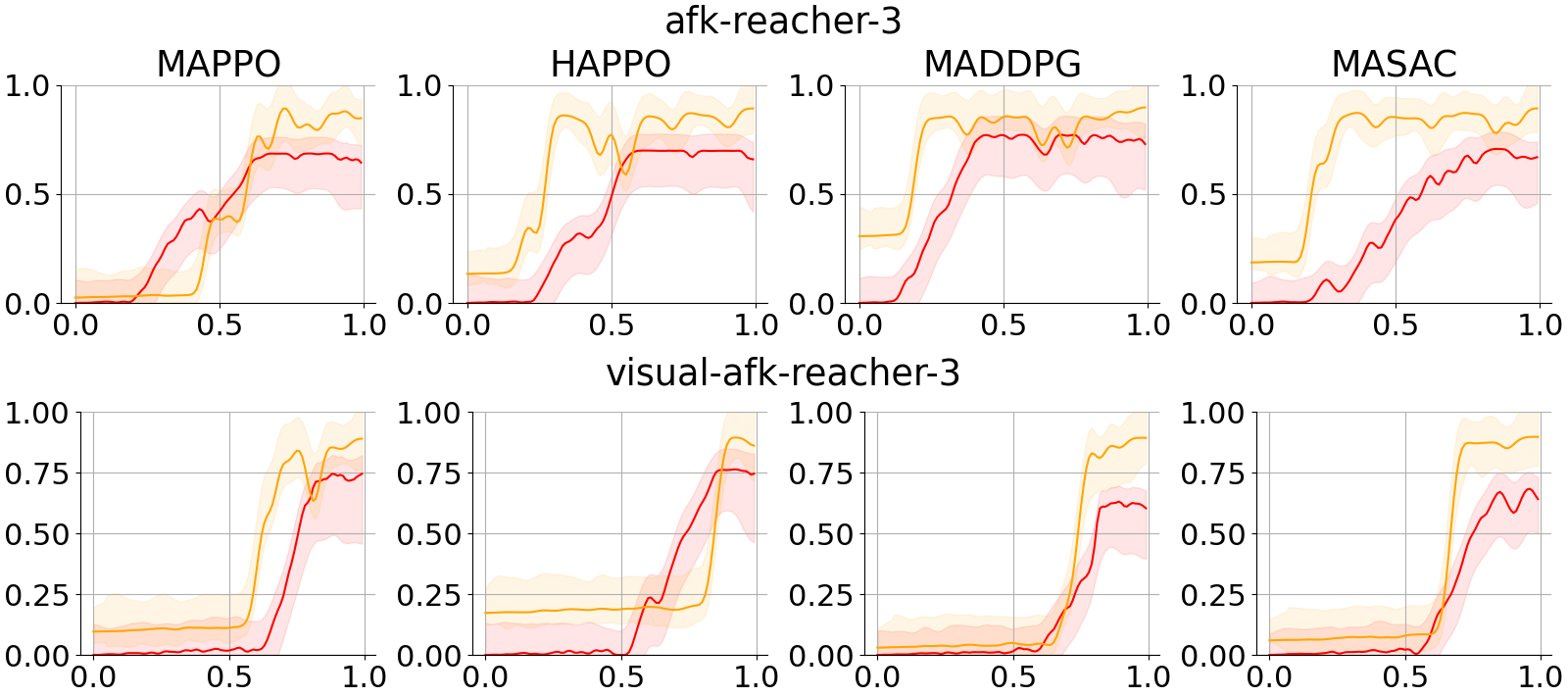} }}}%
    \caption{VMAS/IST Results For Pretraining: These plots compares the performance of MAPO-LSO with (shown in the orange line) and without (shown in the red line) pre-training in each task tested in the VMAS/IST benchmark under a normalized scale. We note that the same hyperparameters are used for both, but they are tuned for MAPO-LSO \textbf{without} pretraining. The format follows Figure \ref{fig:fullist}.}%
    \label{fig:fullpt}%
\end{figure}

\newpage
\section{Full Results: Uncertainty Modeling Experiments} \label{fullnum}
\begin{figure}[!htbp]
    \centering
    \subfloat{{\includegraphics[width=0.45\textwidth]{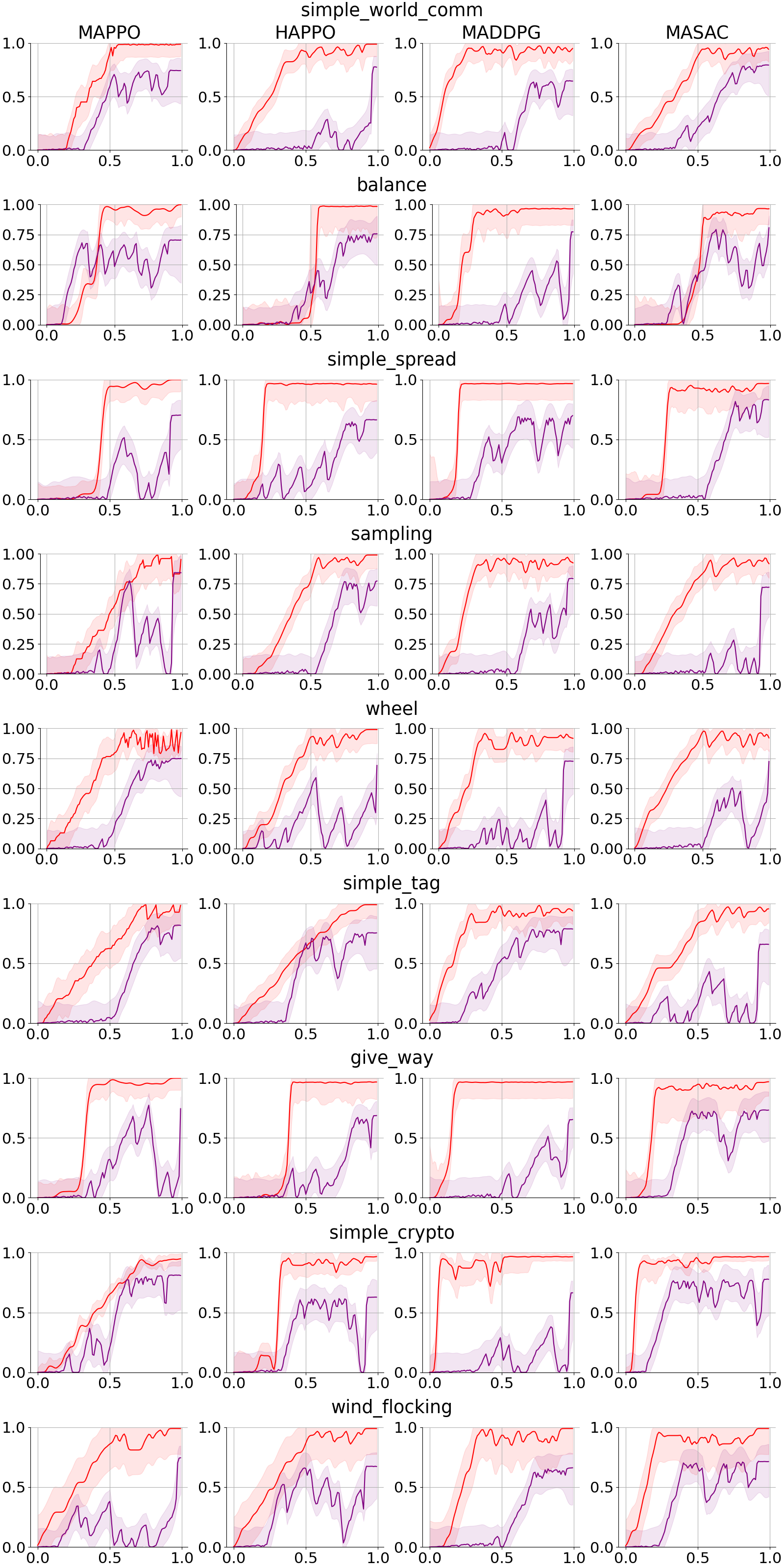} }}%
    \qquad
    \subfloat{{\includegraphics[width=0.45\textwidth]{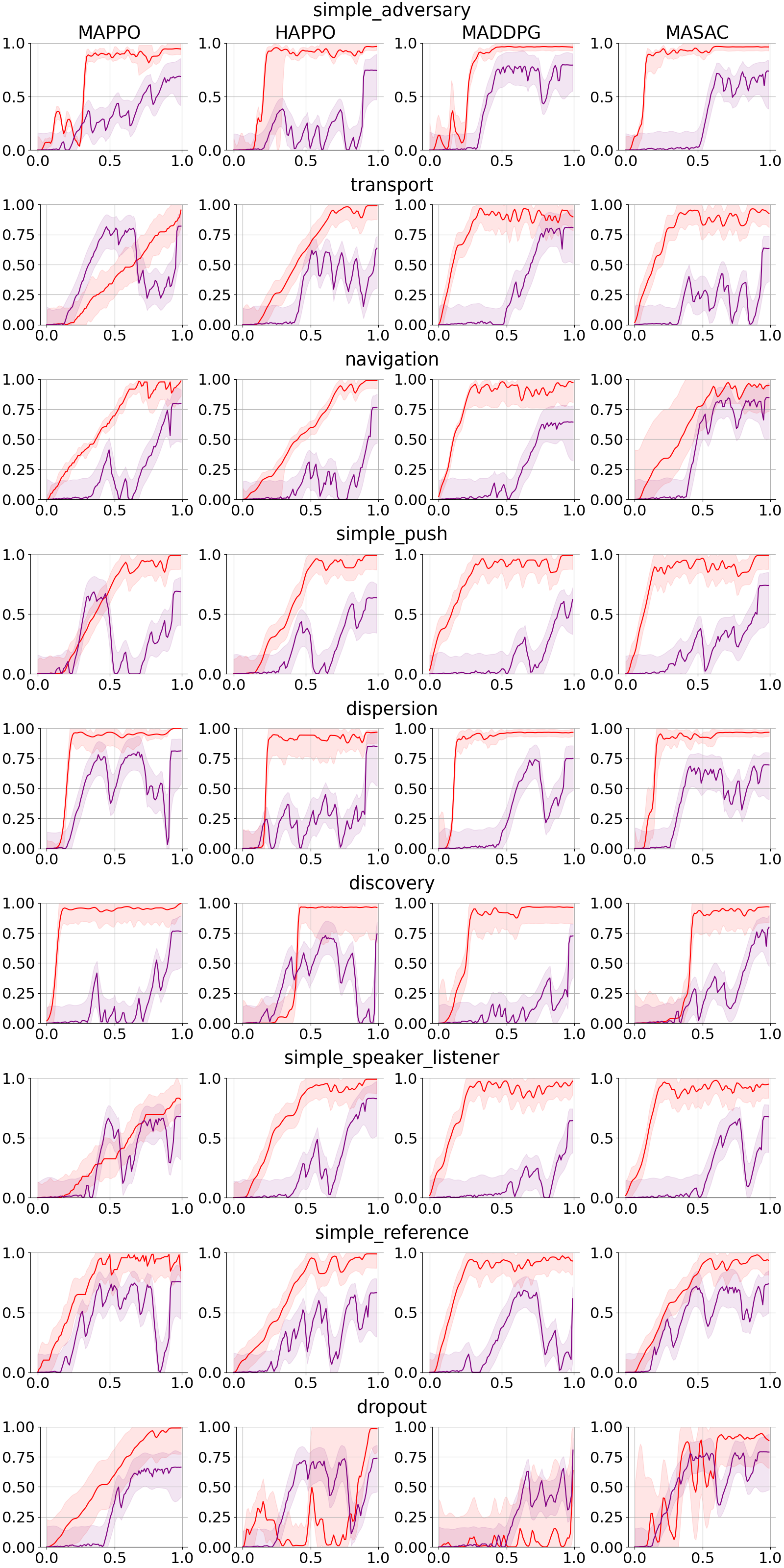} }}%
    \qquad
    \subfloat{{\includegraphics[width=0.45\textwidth]{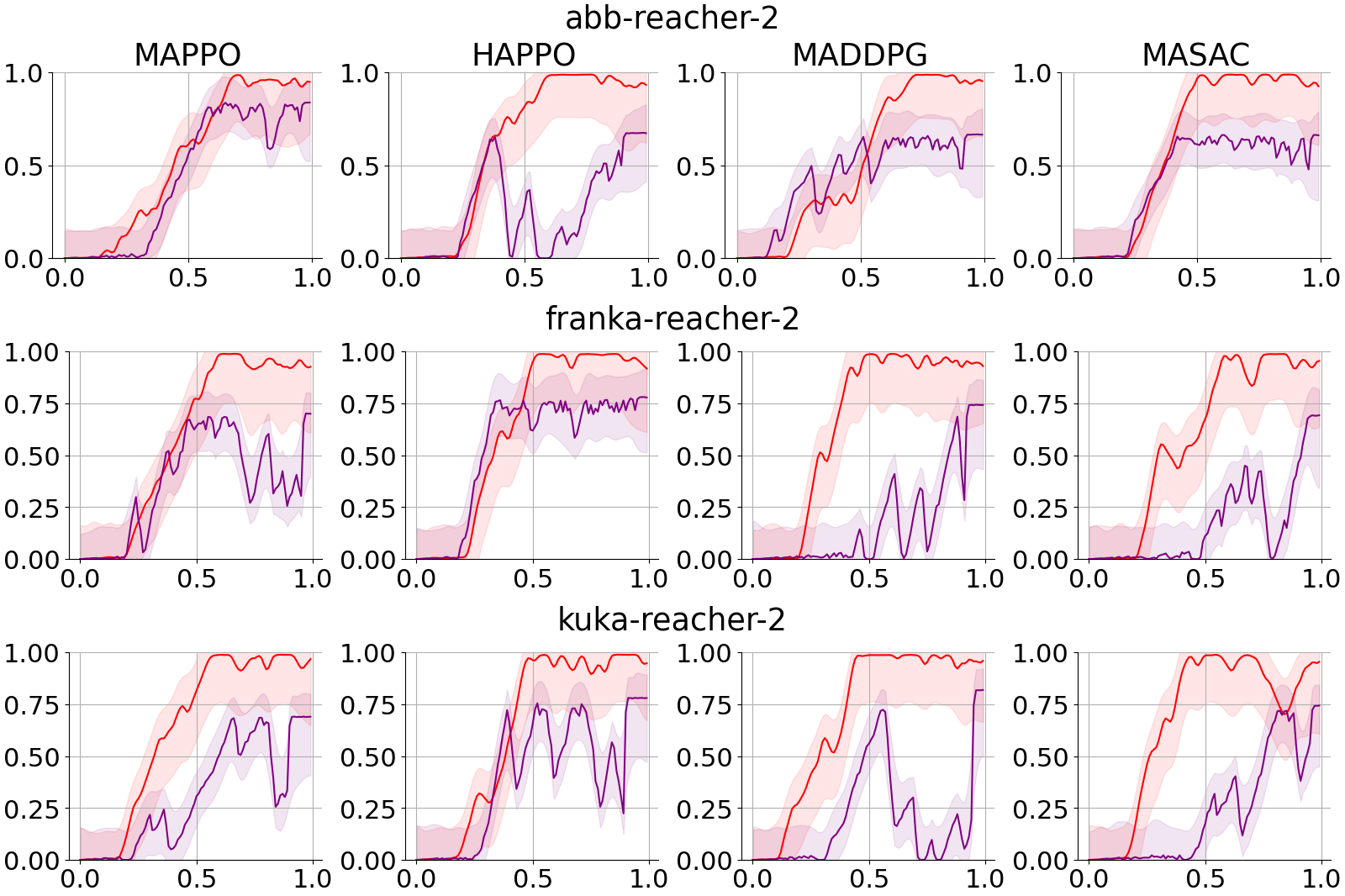} }}%
    \qquad
    \raisebox{4em}{\subfloat{{\includegraphics[width=0.45\textwidth]{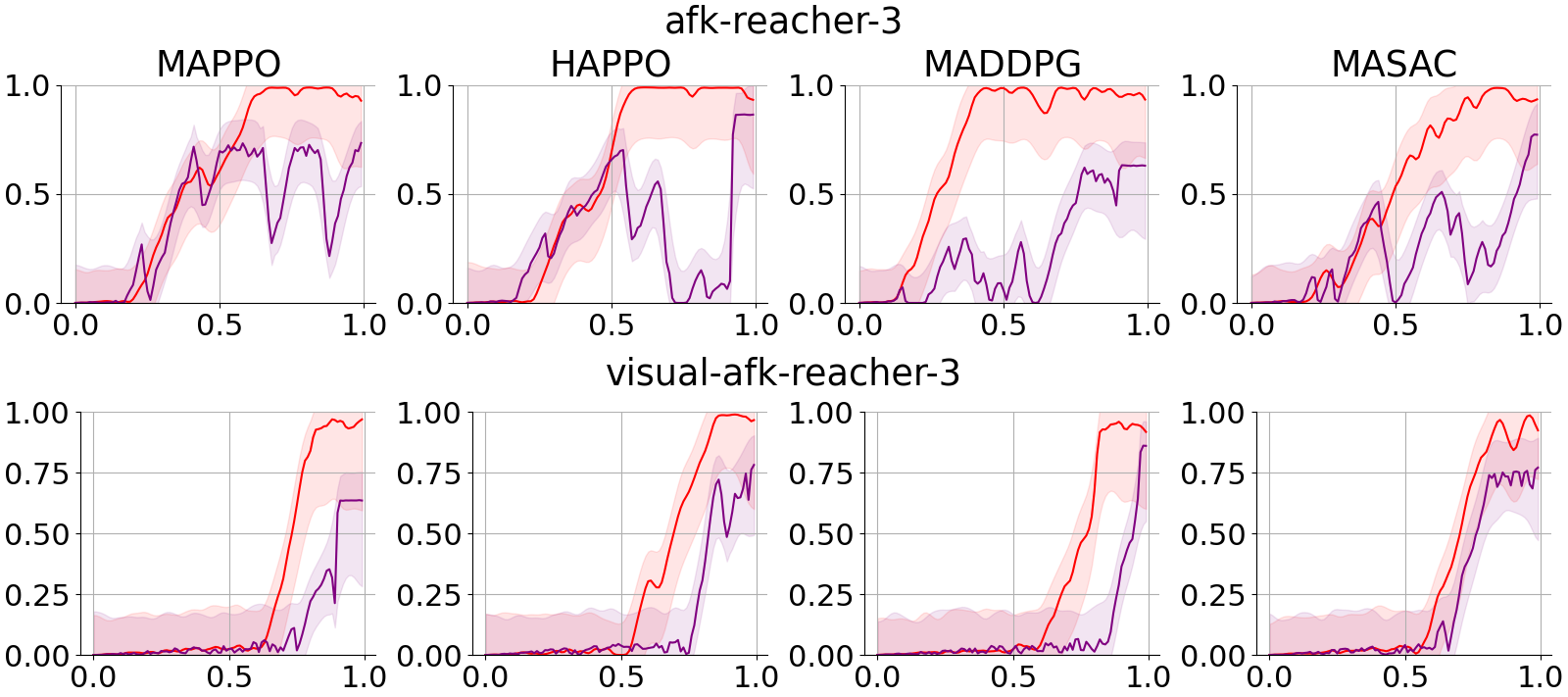} }}}%
    \caption{VMAS/IST Results For Uncertainty Modeling: These plots compares the performance of MAPO-LSO with (shown in the red line) and without (shown in the purple line) uncertainty modeling in each task tested in the VMAS/IST benchmark under a normalized scale. We note that the same hyperparameters are used for both, but they are tuned for MAPO-LSO \textbf{with} uncertainty modeling. The format follows Figure \ref{fig:fullist}.}%
    \label{fig:fullnum}%
\end{figure}

\newpage
\section{Full Results: MAPO-LSO Abalations} \label{fullabalation}
\begin{figure}[!htbp]
    \centering
    \subfloat{{\includegraphics[width=0.45\textwidth]{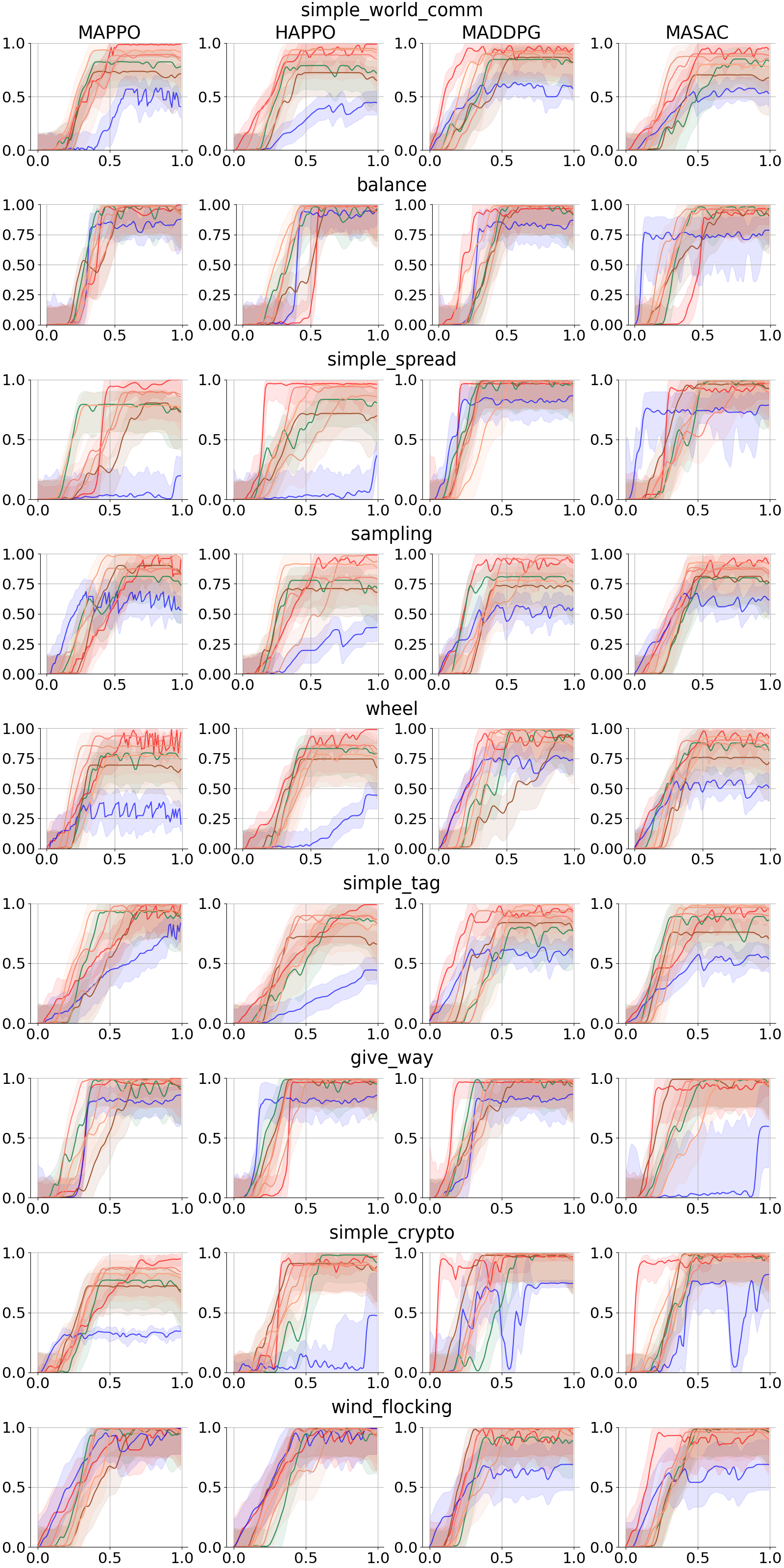} }}%
    \qquad
    \subfloat{{\includegraphics[width=0.45\textwidth]{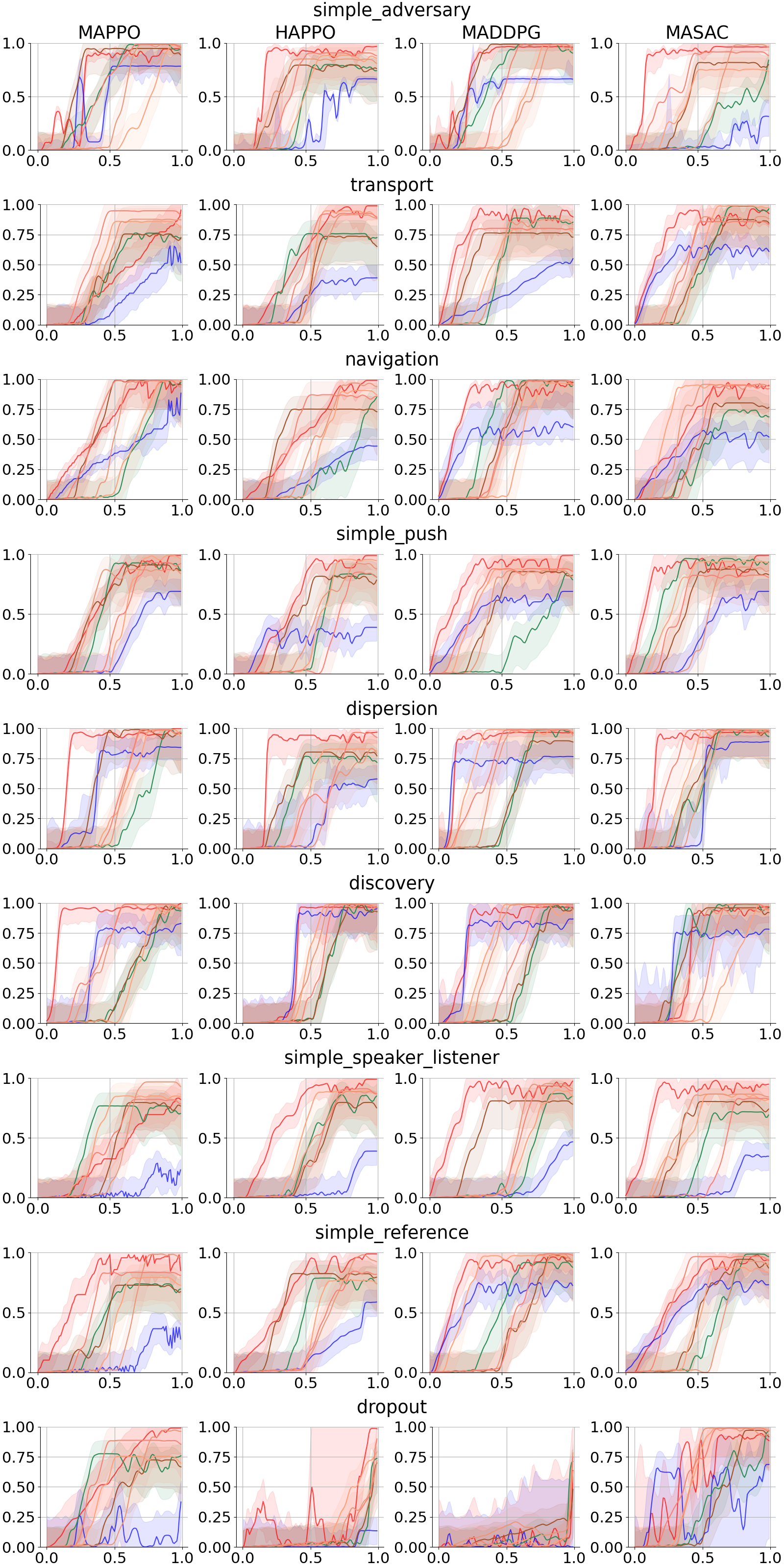} }}%
    \qquad
    \subfloat{{\includegraphics[width=0.45\textwidth]{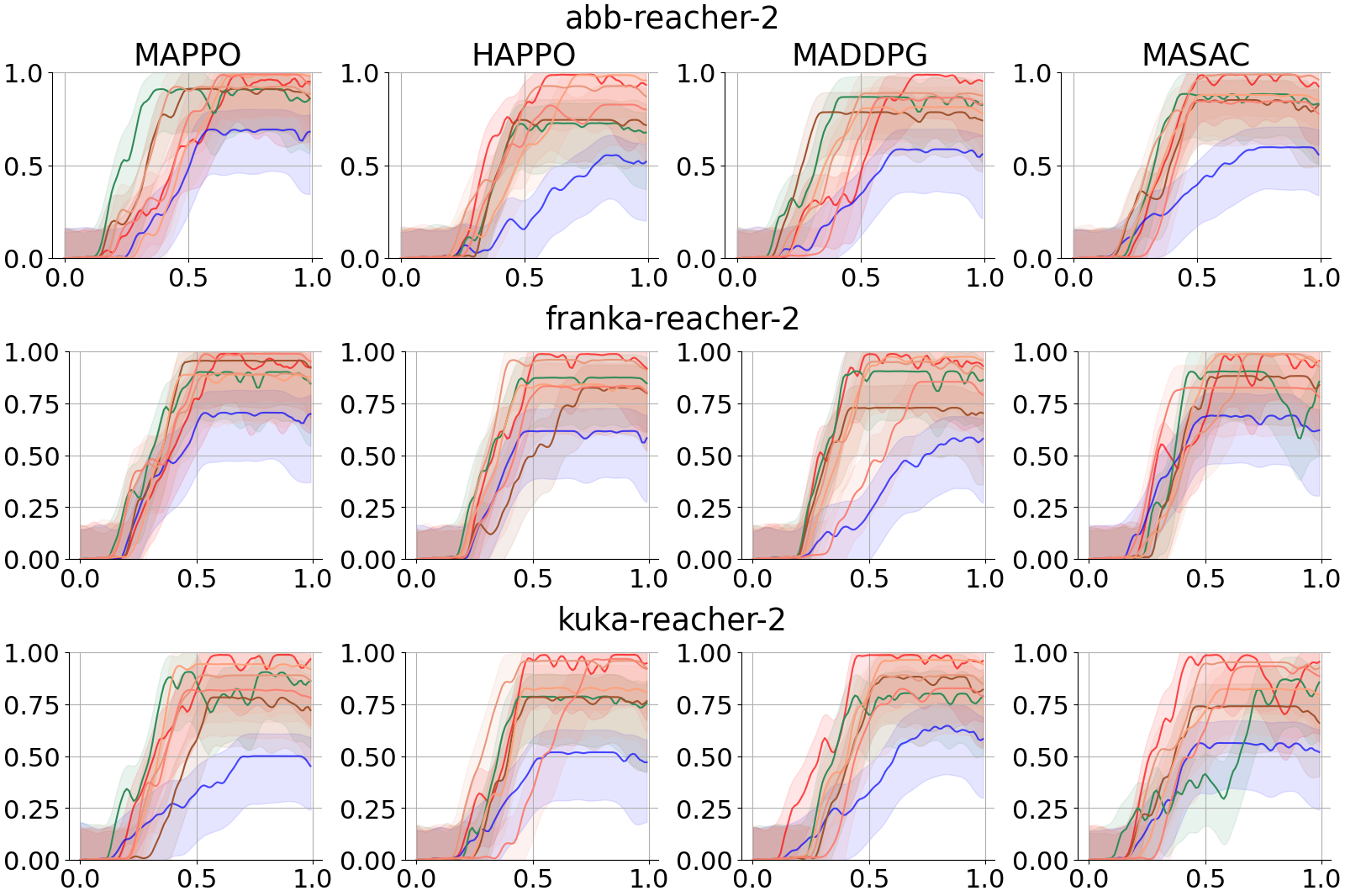} }}%
    \qquad
    \raisebox{4em}{\subfloat{{\includegraphics[width=0.45\textwidth]{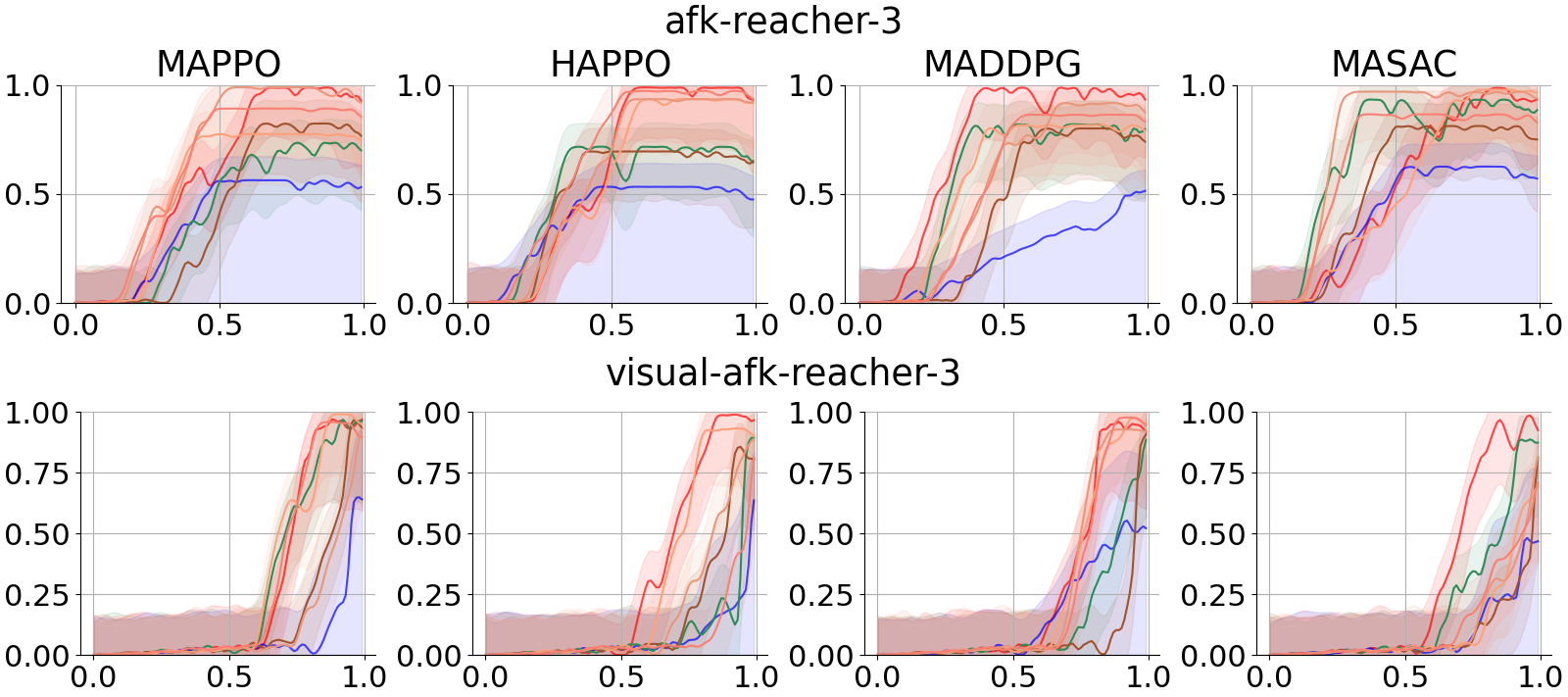} }}}%
    \caption{VMAS Results For MAPO-LSO Abalations: These plots compares the performance of MAPO-LSO with various components missing in each task tested in the VMAS benchmark under a normalized scale. We note that the same hyperparameters are used for all, but they are tuned for MAPO-LSO \textbf{with} all components (LSO) and the vanilla MARL algorithm (no LSO). The format follows Figure \ref{fig:fullist}.}%
    \label{fig:fullablation}%
\end{figure}
\end{document}